\documentclass[final,3p,authoryear,times,sort&compress]{elsarticle}

\usepackage{multirow,setspace,times,amssymb,amsmath,graphicx,color,bm,rotating,subfigure,url,epstopdf}
\usepackage{lineno}
\usepackage{natbib}
\usepackage{array}
\usepackage{diagbox}
\usepackage[colorlinks,linkcolor=red,anchorcolor=blue,citecolor=green]{hyperref}
\usepackage{bigstrut}
\usepackage{threeparttable}
\usepackage[font=small]{caption}
\usepackage{ragged2e}

\usepackage{dcolumn}
\newcolumntype{d}[1]{D{.}{.}{#1}}

\journal{Energy Economics}

\begin{document}

\begin{frontmatter}

\title{{The role of global economic policy uncertainty in predicting crude oil futures volatility: Evidence from a two-factor GARCH-MIDAS model}}
\author[CME]{Peng-Fei Dai}
\author[CME,CCSCA]{Xiong Xiong}
\author[BS,SS,RCE]{Wei-Xing Zhou\corref{cor1}}
\cortext[cor1]{Corresponding author.}
\ead{wxzhou@ecust.edu.cn} %

\address[CME]{College of Management and Economics, Tianjin University, Tianjin 300072, China}
\address[CCSCA]{China Center for Social Computing and Analytics, Tianjin University, Tianjin 300072, China}
\address[BS]{School of Business, East China University of Science and Technology, Shanghai 200237, China}
\address[SS]{Department of Mathematics, East China University of Science and Technology, Shanghai 200237, China}
\address[RCE]{Research Center for Econophysics, East China University of Science and Technology, Shanghai 200237, China}

\begin{abstract}
This paper aims to examine whether the global economic policy uncertainty (GEPU) and uncertainty changes have different impacts on crude oil futures volatility. We establish single-factor and two-factor models under the GARCH-MIDAS framework to investigate the predictive power of GEPU and GEPU changes excluding and including realized volatility. The findings show that the models with rolling-window specification perform better than those with fixed-span specification. For single-factor models, the GEPU index and its changes, as well as realized volatility, are consistent effective factors in predicting the volatility of crude oil futures. Specially, GEPU changes have stronger predictive power than the GEPU index. For two-factor models, GEPU is not an effective forecast factor for the volatility of WTI crude oil futures or Brent crude oil futures. The two-factor model with GEPU changes contains more information and exhibits stronger forecasting ability for crude oil futures market volatility than the single-factor models. The GEPU changes are indeed the main source of long-term volatility of the crude oil futures.
\end{abstract}


\end{frontmatter}


\section{Introduction}
\label{S1:Intro}
The fast-growing commodity markets are attracting the attention of more and more investors and policy makers, since commodity futures broaden the instruments for financial market investment and play an important role in preventing systemic risk. On April 20, 2020, the West Texas Intermediate (WTI) crude oil futures closed at $-\$37.63$ per barrel. This event catches the eyes of the world and produces a profound influence on practitioners and policy makers \citep{Ji-Zhang-Zhao-2020-IRFA}. Therefore, commodity-related research also has practical significance.

A large number of studies show that the commodity futures are valuable sources of diversification investment for investors and portfolio managers \citep{Arouri-Jouini-Nguyen-2011-JIMF,Lucey-Sharma-Vigne-2017-EconM,Klein-2017-FRL,Fazelabdolabadi-2019-FinancInnov}. \cite{Geman-Kharoubi-2008-JBF} explore the diversification effect brought by crude oil futures contracts into a portfolio of stocks. \cite{Nguyen-Sensoy-Sousa-Uddin-2020-EE} study the hedging versus the financialization nature of commodity futures, and find that gold can be seen as a hedge against unfavorable fluctuations in the stock market. \cite{Narayan-Narayan-Zheng-2010-AEn} examine the long-run relationship between oil and gold spot and futures markets. \cite{Hammoudeh-Nguyen-Reboredo-Wen-2014-EMR} provide evidence of low and positive correlations between commodity markets and stock markets and suggest that commodity futures are a desirable asset class for portfolio diversification. Among all the commodities, the price dynamics of crude oil futures and related energy futures play a crucial role in modern global economic and financial systems and our daily life \citep{Jones-Kaul-1996-JF,Sadorsky-1999-EE}. 

The research of crude oil futures can be divided into two strands: price evolution and fluctuation dynamics. In terms of price evolution, scholars have studied the correlation and influence mechanism of crude oil futures and crude oil spot, other commodity futures. In addition, the impact factors of crude oil futures pricing and the influence of crude oil futures market on other financial markets are discussed. \cite{Chang-Lee-2015-EE} and \cite{Holmes-Otero-2019-EE} investigate the correlation and the causality between crude oil futures and spot prices over time using different methods. \cite{Wang-Shao-Kim-2020-CSF} and \cite{Liu-Pan-Yuan-Chen-2019-Energy} detect the correlation of the crude oil futures price with other futures price. \cite{Cheng-Nikitopoulos-Schlogl-2018-JBF} show that the interest rates, the most traditional financial instruments, have influence on crude oil futures prices. \cite{Yan-Irwin-Sanders-2018-EE} and \cite{Ames-Bagnarosa-Matsui-Peters-Shevchenko-2020-EE} discuss other impact factors on crude oil futures prices. By contrast, more scholars focus on the study of the volatility of crude oil futures market \citep{Agnolucci-2009-EE,Kang-Yoon-2013-EE,Ergen-Rizvanoghlu-2016-EE,Liu-Han-Yin-2018-JFutM,Zhang-Ma-Wei-2019-EE,Hasanov-Shaiban-Freedi-2020-EE,Joet-Valerie-2017-EE}. Undoubtedly, exploring the sources of the crude oil futures market volatility is crucial to energy researchers, financial practitioners and policy makers. Our contribution is to expand the literature on the determinants of crude oil futures market volatility.

The determinants of crude oil futures market volatility attract the attention of many scholars \citep{Bakas-Triantafyllou-2019-EE,Liu-Han-Yin-2018-JFutM,Nguyen-Walther-2020-JFc}. For instance, \cite{Bakas-Triantafyllou-2019-EE} study the predictive power of macroeconomic uncertainty on the volatility of agricultural, energy and metals commodity markets. In their paper, the latent macroeconomic uncertainty is constructed by \cite{Jurado-Ludvigson-Ng-2015-AER}. \cite{Liu-Han-Yin-2018-JFutM} investigate the impact of news implied volatility and its sub-component on the volatility of commodity futures. The news implied volatility is introduced by \cite{Manela-Moreira-2017-JFE}, which quantifies the information about uncertainty from newspaper articles. \cite{Bakas-Triantafyllou-2019-EE} and \cite{Liu-Han-Yin-2018-JFutM} also discuss the impact of economic policy uncertainty of the United States on some commodities. \cite{Fang-Chen-Yu-Qian-2018-JFutM} examine whether global economic policy uncertainty contains forecasting information for global gold futures market volatility. To our knowledge, rare literature investigates in depth the influence of global economic policy uncertainty on crude oil futures market volatility.

Nowadays, the ties between different economies are getting stronger and stronger, and the development of world economy is highly integrated \citep{Dai-Xiong-Zhou-2019-PA}. At the same time, the internal factors and external environment that affect economic development are changing over time. Consequently, the global economic policy uncertainty has become a new normal, which is also time-varying. The study on uncertainty has attracted much attention \citep{Bloom-2009-Em,Pastor-Veronesi-2012-JF,Pastor-Veronesi-2013-JFE,Moore-2017-ER,Castelnuovo-Tran-2017-EL}. For instance, \cite{Pastor-Veronesi-2012-JF} and \cite{Pastor-Veronesi-2013-JFE} develop a general equilibrium model to study how policy uncertainty affect stock market. \cite{Baker-Bloom-Davis-2016-QJE} construct a seminal index as the proxy for economic policy uncertainty in the United States and 11 other major economies, which was initially put forward by \cite{Baker-Bloom-Davis-2013-CBRP}. Inspired by \cite{Baker-Bloom-Davis-2016-QJE}, many scholars \citep{Moore-2017-ER,Arbatli-Davis-Ito-Miake-Saito-2017-IMF,Castelnuovo-Tran-2017-EL} propose many other indices for different economies successively using different methods and study the influence of economic policy uncertainty on various financial markets. Since crude oil futures are highly correlated with the global economic environment, as well as the national policy environment, it is crucial to investigate how the global uncertainty related to economic policy affect crude oil futures markets volatility. \cite{Dai-Xiong-Zhou-2020-FRL} construct an index for the aggregate global uncertainty related to economic policy based on the principal component analysis. Hence, using this index as the proxy variable for global economic policy uncertainty and the changes of the index as the proxy variable for the changes of global economic policy uncertainty, we study the determinants of crude oil futures markets volatility. 

There are various methods to model and predict the volatility of the crude oil futures market, among which the GARCH-class models are the most widely used. Most empirical tests require data of the same frequency for the volatility and its potential sources. To overcome this shortfall, \cite{Ghysels-Santa-Clara-Valkanov-2004} and \cite{Ghysels-Arthur-Rossen-2007-EmRev} introduce and re-explore MIDAS regression models, which can deal with time series data sampled at different frequencies. \cite{Engle-Rangel-2008-RFS} propose the spline-GARCH model to combine the macroeconomic causes with low-frequency volatility of equities. In their model, high-frequency return volatility is specified to be the product of a slow-moving component, represented by an exponential spline, and a unit GARCH. \cite{Engle-Ghysels-Sohn-2013-RES} formulate a new class of component models, i.e. GARCH-MIDAS models, which distinguish long-term movement from short-term movement. \cite{Wei-Liu-Lai-Hu-2017-EE} investigate the informative determinant in forecasting crude oil spot market volatility via employing the GARCH-MIDAS model. \cite{Asgharian-Hou-Javed-2013-JFc} utilize the GARCH-MIDAS model to examine the forecasting power of macroeconomic variables on short-term and long-term component of the variance of equity returns. They detect a large group of macroeconomic variables including unexpected inflation, term premium, per capita labour income growth, default premium, unemployment rate, short-term interest rate, and per capita consumption. \cite{Asgharian-Hou-Javed-2013-JFc} augment the model by adding the level and variance of an economic variable to the MIDAS model. Based on the model proposed by \cite{Asgharian-Hou-Javed-2013-JFc}, \cite{Fang-Chen-Yu-Qian-2018-JFutM} investigate whether global economic policy uncertainty contains forecasting information for global gold futures market volatility. There is little literature focusing on the different effects between global economic policy uncertainty and its changes. Our paper contributes to the literature on modelling the influence of global economic policy uncertainty and its changes on crude oil futures volatility.

The remainder of the paper is organized as follows. Section~\ref{S2:Data} describes the data. Section~\ref{S3:Method} presents the models and evaluation methods. In Section~\ref{S4:EmpRes}, the empirical results are reported. Section~\ref{S5:Conclusion} concludes the paper.

\section{Data description}
\label{S2:Data}

\begin{figure}[htbp]
  \centering
  \includegraphics[width=0.49\linewidth]{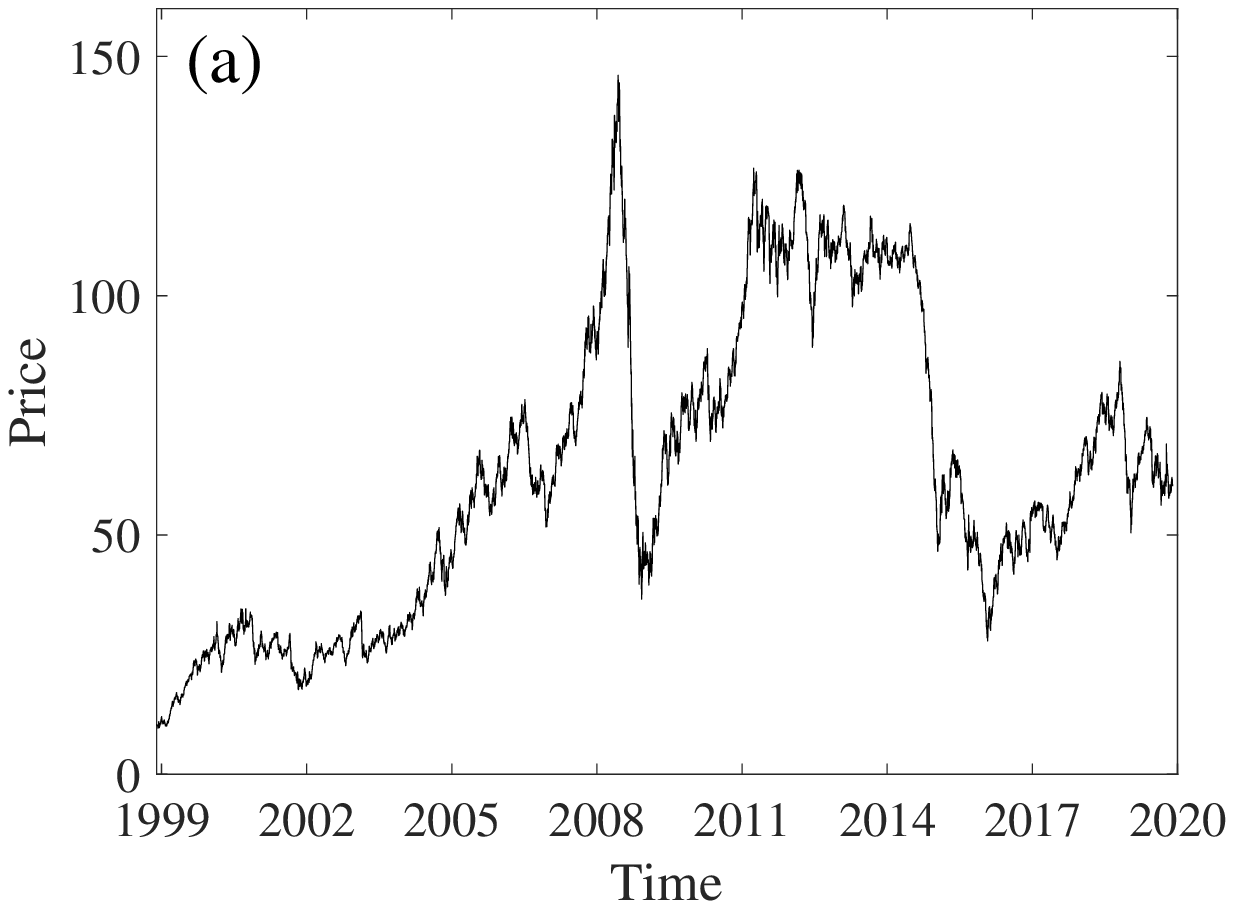}
  \includegraphics[width=0.49\linewidth]{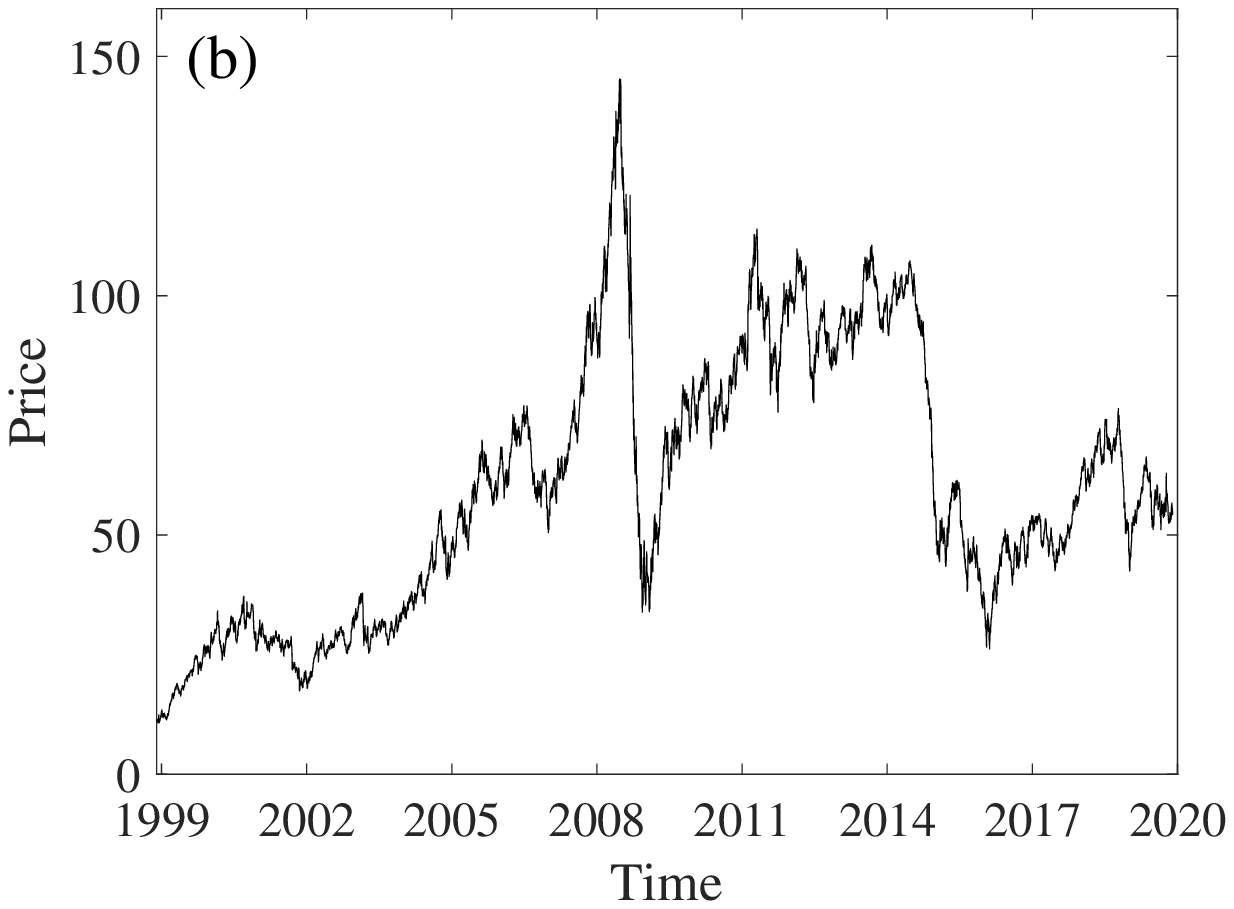}
  \caption{The evolutionary trajectories of Brent crude oil futures price (a) and WTI crude oil futures price (b). The sample period of the two crude oil futures is from 1 December 1998 to 31 October 2019.}
\label{Fig:CrudeOilVolatility:Price}
\end{figure}

The dominating global crude oil futures markets are the New York Mercantile Exchange (NYMEM) in the United States and the Intercontinental Exchange (ICE) in the United Kingdom. The two most important pricing benchmarks for the global oil market are West Texas Intermediate (WTI) crude oil futures contracts traded on the NYMEM and Brent crude oil futures contracts traded on the ICE. In this work, we choose the two commodity futures to represent the crude oil futures market. And their ``contract 1'' are selected for subsequent analyses. We retrieve the daily prices of the two crude oil futures from the web site of the U.S. Energy Information Administration and the prices are in dollars per barrel. In order to match the data of the GEPU index, the samples are from 1 December 1998 to 31 October 2019. Figure~\ref{Fig:CrudeOilVolatility:Price} illustrates the evolutionary price trajectories of the two commodity futures. The daily returns of crude oil futures are calculated as follows
\begin{equation}
  r_t=\ln\left( \frac{p_t}{p_{t-1}}\right),
  \label{Eq:rt}
\end{equation}
where $t$ is in units of trading days.

\begin{figure}[htbp]
\centering
\includegraphics[width=0.49\linewidth]{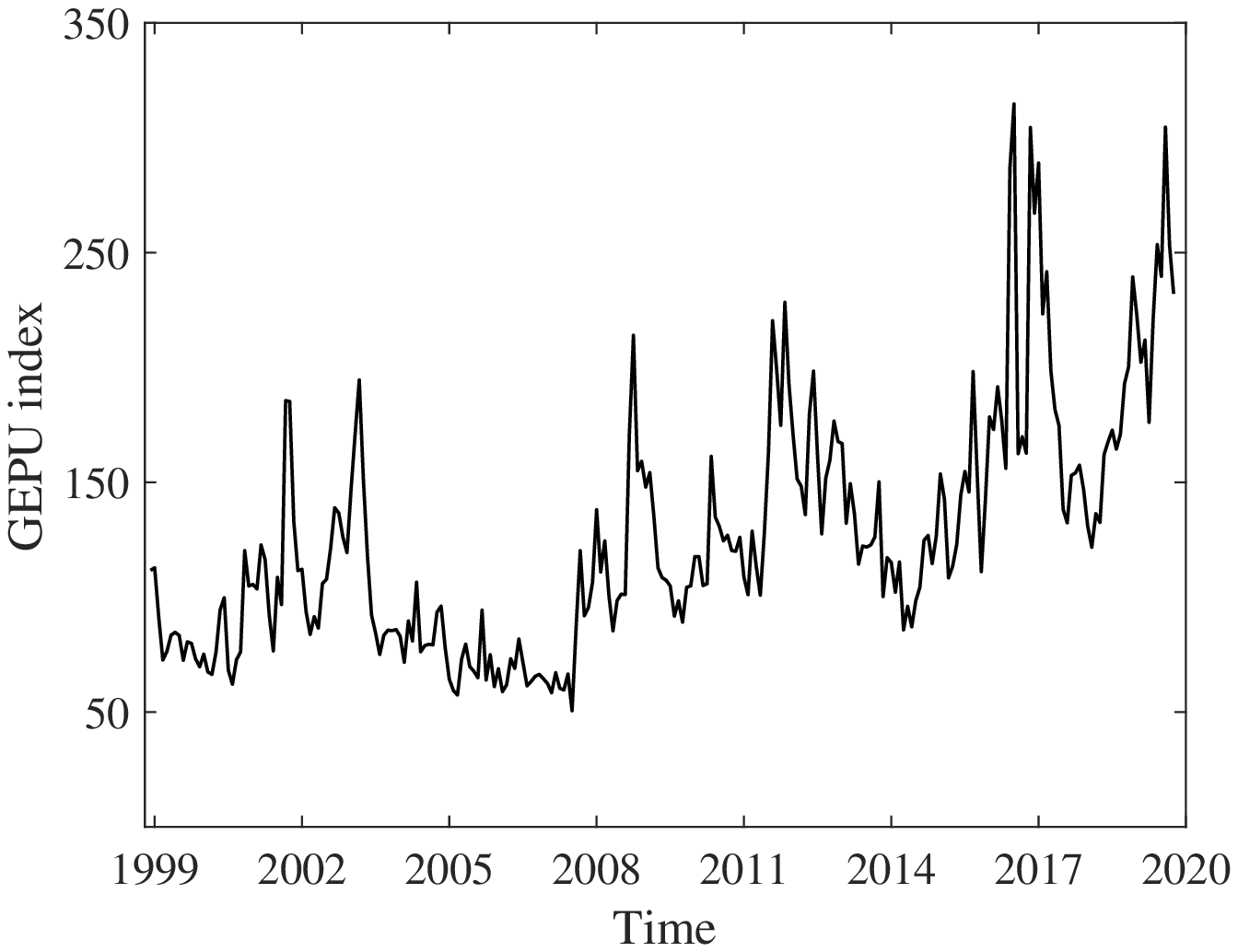}
\includegraphics[width=0.49\linewidth]{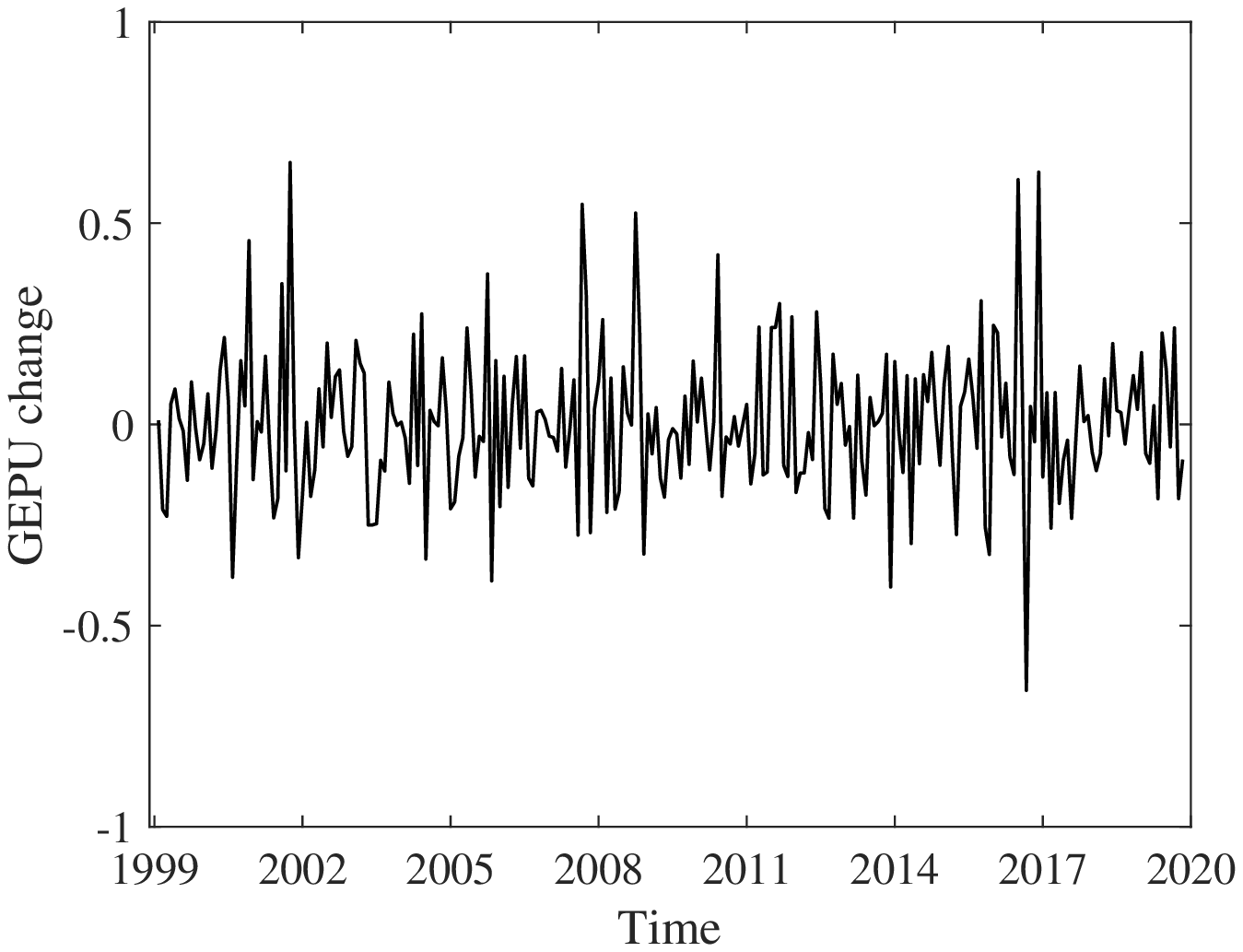}
\caption{The time evolution of the monthly GEPU index and its changes. The sample period of the GEPU index is from December 1998 to October 2019, while the sample of GEPU change covers the period from January 1999 to October 2019.}
\label{Fig:CrudeOilVolatility:GEPU}
\end{figure}

\cite{Dai-Xiong-Zhou-2020-FRL} construct a new global economic policy uncertainty index (GEPU) based on the principal component analysis, which performs comparatively and slightly better in some situations as the GDP-weighted GEPU of \cite{Davis-2016-NBER}. The monthly GEPU index between December 1998 and October 2019 is calculated for our subsequent analysis. In order to carry out the calculation, we select 21 EPU indices representing various economies' economic policy uncertainty\footnote{The EPU indices of the 21 economies are publicly available at \url{http://www.policyuncertainty.com}.}. The 21 economies are Australia, Brazil, Canada, Chile, China, Colombia, France, Germany, Greece, India, Ireland, Italy, Japan, South Korea, Mexico, the Netherlands, Russia, Spain, Sweden, the United Kingdom, and the United States. 
In addition to the GEPU, we pay attention to the corresponding uncertainty change which is named GEPU change. The GEPU changes are calculated as follows
\begin{equation}
  \Delta{GEPU_m}=\ln\left(\frac{GEPU_m}{GEPU_{m-1}}\right),
  \label{Eq:GEPU}
\end{equation}
where $m$ is in units of months.
Figure~\ref{Fig:CrudeOilVolatility:GEPU} illustrates the time series of the monthly GEPU inedx and its changes. Comparing Fig.~\ref{Fig:CrudeOilVolatility:GEPU} and Fig.~\ref{Fig:CrudeOilVolatility:Price}, we see that the GEPU index rose rapidly around the global financial crisis in 2008-2009, while both the Brent and WTI crude oil futures prices plummeted during the period. After the global financial crisis, the crude oil futures prices recovered and stabilized, whereas the GEPU index maintained at a relatively high level. 

Table~\ref{TB:CrudeOilVolatility:DataStat} presents the summary statistics of the four time series, where the data frequency, mean, minimum, maximum, standard error, skewness and kurtosis are reported. The GEPU index and its changes are monthly, whose sampling frequencies are lower than the daily crude oil futures returns. All the means of the time series are close to zero except for the GEPU index whose mean is 152.81. The distributions of the GEPU index and its changes are positively skewed and leptokurtic, while the two crude oil futures returns' distributions are negatively skewed and leptokurtic. In addition, We verify whether each time series is stationary using the augmented Dickey-Fuller (ADF) test. We find that the the GEPU index and its changes, as well as the two crude oil futures returns time series, are stationary. The result of the stationary test for the GEPU index is in line with that for the GDP-weighted GEPU \citep{Fang-Bouri-Gupta-Roubaud-2019-IRFA}. Thus, all of the time series can be modelled directly. 

\begin{table}[htp]
\centering
\setlength{\abovecaptionskip}{0pt}
\setlength{\belowcaptionskip}{10pt}
\caption{Summary statistics of the crude oil futures returns, the GEPU index, and the GEPU changes.}
\smallskip
\label{TB:CrudeOilVolatility:DataStat}
\begin{threeparttable}  
\begin{center}
\setlength{\tabcolsep}{2.55mm}{ 
\begin{tabular}{cccccccccc}
 \hline \hline
 Variable      &  Obs.  &   Freq.  &    Mean     &    Min.    &  Max.   &   Std. dev.  & Skew.     &  Kurt.  &     ADF       \\ \hline
 $r$ (Brent)       & 5258  &  daily   &  3.27E-04  &   $-0.14$  &  0.14   &   0.02       & $-0.11$   &   6.02  & $-77.03^{***}$ \\
 $r$ (WTI)         & 5258  &  daily   &  2.99E-04  &   $-0.17$  &  0.16   &   0.02       & $-0.14$   &   7.04  & $-75.29^{***}$ \\ 
 $GEPU$      & 251   &  monthly &  152.81     &   50.55    &  314.70 &   52.03      & 1.14      &   4.33  &     $-3.43^{***}$  \\
 $\Delta{GEPU}$    & 250   &  monthly &  2.90E-03  &   $-0.66$  &  0.65   &   0.18       & 0.46      &   4.70  & $-18.07^{***}$ \\
\hline
\end{tabular}
}
\end{center}
\footnotesize       
\justifying Note: ADF represents the statistics calculated from the unit root test, and $^{***}$, $^{**}$, $^{*}$ denote the null hypothesis is rejected at $1\%$, $5\%$, $10\%$ statistical significance level respectively.
\end{threeparttable}
\end{table}

\section{Empirical methodology}
\label{S3:Method}

We utilize the GARCH-MIDAS model to investigate the effects of economic policy uncertainty on the daily price volatility of energy futures markets. \cite{Engle-Ghysels-Sohn-2013-RES} propose the GARCH-MIDAS model to study the contribution of macroeconomic variables to stock volatility. They decompose the volatility of low-frequency time series into two components: short-term volatility and long-term volatility. The long-run volatility is determined by low-frequency macroeconomic factors, while the short-run volatility depends on the dynamics of the high-frequency time series itself. Following this thread, we introduce the GEPU index and its changes into the GARCH-MIDAS model so as to explore the effects of economic policy uncertainty on the long-term volatility of crude oil futures. Besides the macroeconomic variable, the contribution of the realized volatility is also considered. To carry out better the research, two models are specified, the single-factor model and two-factor model.

\subsection{Single-factor model}
\label{S3:Method:SinfactorMod}

The GARCH-MIDAS model could be formally expressed as follows. The return on day $i$ in period $t$ (which may be a week, a month, a quarter or longer) follows the following process:
\begin{equation}
  r_{i,t}=E_{i-1,t}\left(r_{i,t}\right)+\sqrt{\tau_tg_{i,t}}\varepsilon_{i,t}, \; {\forall}i=1,\ldots,N_t 
\label{Eq:rit}
\end{equation}
and
\begin{equation}
  E_{i-1,t}\left(r_{i,t}\right)=\mu,~~~~ \varepsilon_{i,t}\mid\Phi_{i-1,t}\sim N\left(0,1\right),
\end{equation}
where $N_t$ is the number of trading days in each period, $\Phi_{i-1,t}$ is the information set up to day $i-1$, and $\varepsilon_{i,t}$ is the innovation term. We set $\mu$ as a constant since the mean daily return of crude oil futures is quite small. Eq.~(\ref{Eq:rit}) implies the volatility of the return is decomposed into two parts: one is a short-term volatility component represented by $g_{i,t}$ , and the other is a long-term volatility component represented by $\tau_t$.
    
The dynamics of the short-term volatility component $g_{i,t}$ is assumed to be a daily GARCH $\left(1,1\right)$ process:
\begin{equation}
  g_{i,t}=(1-\alpha-\beta)+\alpha\frac{(r_{i-1,t}-\mu)^2}{\tau_t}+\beta{g_{i-1,t}},
  \label{Eq:git}
\end{equation}
where $\alpha>0$,~$\beta>0$ and $\alpha+\beta<1$, while the long-term volatility component $\tau_t$ is specified as smoothed realized volatility in the spirit of the MIDAS regression:
\begin{equation}
  \tau_t=m+\theta\sum_{k=1}^K\phi_k(\omega_1,\omega_2)RV_{t-k},
  \label{Eq:tau:RV:fixspan}
\end{equation}
where
\begin{equation}
RV_t=\sum_{i=1}^{N_t}r_{i,t}^2
\label{Eq:RV:fixspan}
\end{equation}
is the realized volatility in period $t$, $K$ is the number of periods over which we smooth the realized volatility, $m$ is the intercept, and $\theta$ is the slope denoting the impact of realized volatility on long-term volatility. Following \cite{Engle-Ghysels-Sohn-2013-RES}, the weighting scheme in Eq.~(\ref{Eq:tau:RV:fixspan}) is assigned by a two-parameter Beta polynomial:
\begin{equation}
\phi_k(\omega_1,\omega_2)
=\frac{k^{\omega_1-1}(K-k)^{\omega_2-1}}{\sum_{j=1}^Kj^{\omega_1-1}(K-j)^{\omega_2-1}}.
\label{Eq:weight}
\end{equation}
Eqs.~(\ref{Eq:rit}-\ref{Eq:weight}) constitute the single-factor model pertaining to realized volatility under the GARCH-MIDAS framework (Model I) . The realized volatility figured out from Eq.~(\ref{Eq:RV:fixspan}) is fixed in period $t$, where we set the period as a month because the GEPU index and its changes are monthly data. 

We further consider the single-factor model with the realized volatility in rolling-window specification. In this case, Eq.~(\ref{Eq:rit}), Eq.~(\ref{Eq:git}) and Eq.~(\ref{Eq:weight}) are almost unchanged, and only expressions of Eq.~(\ref{Eq:tau:RV:fixspan}) and Eq.~(\ref{Eq:RV:fixspan}) are modified. The rolling-window realized volatility is expressed as
\begin{equation}
RV_i^{({\rm{rw}})}=\sum_{j=1}^{N'}r_{i-j}^2,
\label{Eq:RV:RW}
\end{equation}
where $r_{i-j}$ denotes the backward rolling daily returns across various months, and $N'$ is the number of trading days in one month. For simplicity, we pose $N'=22$ in our models. Thereupon, the long-term volatility process is redefined accordingly as follows:
\begin{equation}
\tau_i^{({\rm{rw}})}=m^{({\rm{rw}})}+\theta^{({\rm{rw}})}\sum_{k=1}^K\phi_k(\omega_1,\omega_2)RV_{i-k}^{({\rm{rw}})}.
\label{Eq:tau:RV:RW}
\end{equation}
Finally, we adjust the low-frequency variable in Eq.~(\ref{Eq:rit}) and Eq.~(\ref{Eq:git}) into daily variable. The adjusted Eq.~(\ref{Eq:rit}) and Eq.~(\ref{Eq:git}), together with Eq.~(\ref{Eq:RV:RW}) and Eq.~(\ref{Eq:tau:RV:RW}) form the single-factor model with rolling-window realized volatility (Model II).

Next, we turn to the models that incorporate the GEPU index and its changes directly. We consider the fixed-span specification where the value of long-term volatility is the same on any day in a month and the rolling-window specification which has time-varying long-term volatility in a month. For the fixed-span model with the GEPU index or its changes, the long-term volatility term $\tau$ is expressed as follows:
\begin{equation}
\tau_t=m_l+\theta_l\sum_{k=1}^K\phi_k\left(\omega_1,\omega_2\right)GEPU_{t-k},
\label{Eq:tau:GEPU:fixspan}
\end{equation}
and
\begin{equation}
\tau_t=m_{\Delta}+\theta_{\Delta}\sum_{k=1}^K\phi_k\left(\omega_1,\omega_2\right)\Delta{GEPU}_{t-k},
\label{Eq:tau:GEPUchange:fixspan} 
\end{equation}
where $\Delta{GEPU}$ denotes the GEPU changes. The single-factor model with the GEPU index (Model III) is consisted of Eq.~(\ref{Eq:rit}), Eq.~(\ref{Eq:git}) and Eq.~(\ref{Eq:tau:GEPU:fixspan}). Eq.~(\ref{Eq:tau:GEPUchange:fixspan}) with Eq.~(\ref{Eq:rit}) and Eq.~(\ref{Eq:git}) form the single-factor model with the GEPU changes (Model IV). Turning to the rolling-window setting, the long-term volatility term $\tau$ related to the GEPU index or its changes is specified as:
\begin{equation}
\tau_i^{({\rm{rw}})}=m_l^{({\rm{rw}})}+\theta_l^{({\rm{rw}})}\sum_{k=1}^K\phi_k\left(\omega_1,\omega_2\right)GEPU_{i-k}^{({\rm{rw}})},
\label{Eq:tau:GEPU:RW}   
\end{equation}
and
\begin{equation}
\tau_i^{({\rm{rw}})}=m_{\Delta}^{({\rm{rw}})}+\theta_{\Delta}^{({\rm{rw}})}\sum_{k=1}^K\phi_k\left(\omega_1,\omega_2\right)\Delta {GEPU}_{i-k}^{({\rm{rw}})}.
\label{Eq:tau:GEPUchange:RW}  
\end{equation}
The rolling-window specifications are calculated by the trading day, hence the long-term volatility is not an unchanged value in any month. We adjust Eq.~(\ref{Eq:rit}) and Eq.~(\ref{Eq:git}) into low-frequency expressions, which together with Eq.~(\ref{Eq:tau:GEPU:RW}) or Eq~(\ref{Eq:tau:GEPUchange:RW}) form the single-factor model with the GEPU index (Model V) or its changes (Model VI) respectively. All weighting schemes $\phi_k\left(\omega_1,\omega_2\right)$ appeared in the specifications above have the same definition as presented in Eq.~(\ref{Eq:weight}).

\subsection{Two-factor model}
\label{S3:Method:TwofactorMod}

Apart from the influence of realized volatility itself, how economic policy uncertainty affects the volatility of crude oil futures is worth studying. In order to find out the answer, we integrate the GEPU index or its changes with the realized volatility and get a new presentation of long-term volatility with fixed span:
\begin{equation}
  \tau_t = m +\theta_{RV}\sum_{k=1}^K\phi_k\left(\omega_1,\omega_2\right)RV_{t-k}
  + \theta_{MV}\sum_{k=1}^K\phi_k(\omega_1,\omega_2)M_{t-k}.
\label{Eq:tau:RV+M:fixspan} 
\end{equation}
Accordingly, the rolling-window long-term volatility case is specified as:
\begin{equation}
  \tau_i^{({\rm{rw}})} = m^{({\rm{rw}})}+\theta_{RV}^{({\rm{rw}})}\sum_{k=1}^K\phi_k\left(\omega_1,\omega_2\right)RV_{i-k}^{({\rm{rw}})}
  + \theta_{MV}^{({\rm{rw}})}\sum_{k=1}^K\phi_k\left(\omega_1,\omega_2\right)M_{i-k}^{({\rm{rw}})}.
\label{Eq:tau:RV+M:RW}  
\end{equation}
The variable $M$ in Eq.~(\ref{Eq:tau:RV+M:fixspan}) and Eq.~(\ref{Eq:tau:RV+M:RW}) represents the GEPU index or its changes. As for $M^{\rm{(rw)}}$, firstly we make the macroeconomic variable $M$ to be the daily index through copying the corresponding monthly value to each day in that month. Then we carry out the rolling-window calculation in Eq.~(\ref{Eq:tau:RV+M:RW}). We adopt the same lag order for realized volatility and macroeconomic variables in both fixed-span and rolling-window versions. The weighting coefficients have the same definition as in Eq.~(\ref{Eq:weight}). Eq.~(\ref{Eq:tau:RV+M:fixspan}) with Eq.~(\ref{Eq:rit}) and Eq.~\ref{Eq:git} constitute the two-factor model with fixed span (GEPU index: Model VII and GEPU changes: Model VIII), and Eq.~(\ref{Eq:tau:RV+M:RW}) with adjusted Eq.~(\ref{Eq:rit}) and Eq.~(\ref{Eq:git}) constitute the two-factor model with rolling window (GEPU index: Model IX and GEPU changes: Model X).  

Finally, the total conditional variance in Eq.~(\ref{Eq:rit}) is shown as follows:
\begin{equation}
\sigma_{i,t}^2=\tau_t\cdot{g_{i,t}}.
\label{Eq:sigma}
\end{equation}

\subsection{Model calibration and evaluation}
\label{S3:Method:ModEst&Eva}

We calibrate the models using full sample data and investigate the explanatory ability of the models. In order to implement model evaluation, we start with in-sample calibrations. We estimate the models using a calibration window and then use the estimated parameters to make out-of-sample variance prediction. We choose a thirteen-year calibration window for both WTI oil and Brent oil, then data lagged three years before the calibration window are needed to compute the historical realized volatility and economic policy uncertainty. To evaluate the variance prediction of a specific model, we use two popular loss functions, root mean squared error (RMSE) and root mean absolute error (RMAE), defined as follows: 
\begin{equation}
  RMSE = \sqrt{\dfrac{1}{S}\sum_{s=1}^S\left(\sigma_{s+1}^2-E_s\left(\sigma_{s+1}^2\right)\right)^2}
\end{equation}
and
\begin{equation}
  RMAE = \sqrt{\dfrac{1}{S}\sum_{s=1}^S\left|\sigma_{s+1}^2-E_s\left(\sigma_{s+1}^2\right)\right|},
\end{equation}
where $\sigma_{s+1}^2$ is the actual daily total variance on day $s+1$, $E_s\left(\sigma_{s+1}^2\right)$ is the predicated daily total variance for day $s+1$, and $S$ is the length of prediction interval. 

In order to further verify the quality of the two models, we will conduct robustness test with computing two more loss functions, root mean squared deviation (RMSD) and root mean absolute deviation (RMAD), defined as follows:
\begin{equation}
  RMSD = \sqrt{\dfrac{1}{S}\sum_{s=1}^S\left(\sigma_{s+1}-E_t\left(\sigma_{s+1}\right)\right)^2}
\end{equation}
and
\begin{equation}
  RMAD = \sqrt{\dfrac{1}{S}\sum_{s=1}^S\left|\sigma_{s+1}-E_s\left(\sigma_{s+1}\right)\right|}.
\end{equation}

Finally, for the sake of comparing the predictive accuracy of two competing models, the DM test proposed by \cite{Diebold-Mariano-2002-JBES} is adopted:
\begin{equation}
  DM = \frac{\bar{D}}{\sqrt{var\left(D_s\right)}}\sim{N\left(0,1\right)}
\end{equation}
\begin{equation}
  D_s = E_{{\rm{A}},s}^2 - E_{{\rm{B}},s}^2
\end{equation}
where $E_{{\rm{A}},s}$ and $E_{{\rm{B}},s}$ are the forecast errors of two competing models A and B respectively, $\bar{D}$ is the mean of the time series $D_s$, and ${var\left(D_s\right)}$ is the variance of $D_s$.

\section{Empirical results}
\label{S4:EmpRes}

In this section, we present the calibration results of all the single-factor models in Section~\ref{ModEst:SinfactorMod} and two-factor models in Section~\ref{ModEst:TwofactorMod} from the full sample and investigate the explanatory ability of the models to the long-term volatility of crude oil futures. Next, in Section~\ref{ModEva}, we consider the in-sample estimation and make evaluation for the predictive performance of the models using the out-of-sample prediction errors. Concerning model calibration, we set $\omega_1=1$, following \cite{Engle-Ghysels-Sohn-2013-RES} and \cite{Asgharian-Hou-Javed-2013-JFc}.

\subsection{Calibration of single-factor models}
\label{ModEst:SinfactorMod}

In order to test the explanatory power of the models precisely, we estimate the parameters of the models using full sample data. Panel A of Table~\ref{TB:OilFutGEPU:Calibration:SinFactor} provides the parameter estimates of the single-factor models with fixed-span $RV$ and rolling-window $RV^{(\rm{rw})}$. In Panel A of Table~\ref{TB:OilFutGEPU:Calibration:SinFactor}, all the $\alpha$'s and $\beta$'s values of the two crude oil futures are significantly different from 0 at $1\%$ level and the sum of $\alpha$ and $\beta$ for each commodity futures is less than and close to 1, which implies the short-term volatility of returns in crude oil futures market has clustering features. The parameter $\theta$ reflects how realized volatility affects long-term volatility of crude oil futures return. All the estimates of $\theta$ shown in Panel A of Table~\ref{TB:OilFutGEPU:Calibration:SinFactor} are significantly positive at the $1\%$ level, which means the realized volatility has a positive influence on the long-term volatility of crude oil futures return. Comparing the differences between $\theta$ in Model I and Model II, we find that $\theta$ in Model I is less than that in Model II for both commodities. The realized volatility with rolling-window pattern has greater impact on long-term volatility of crude oil futures return. The value of BIC in Model II is smaller than that in Model I, which is the evidence that realized volatility using rolling-window expression in Eq.~(\ref{Eq:RV:RW}) is a better explanatory factor.

\begin{table}[htp]
\centering
\setlength{\abovecaptionskip}{0pt}
\setlength{\belowcaptionskip}{10pt}
\caption{Parameter calibration of single-factor models}
\smallskip
\label{TB:OilFutGEPU:Calibration:SinFactor}
\resizebox{\textwidth}{!}{
\begin{tabular}{llcccccccc}
  \hline \hline
 Commodity     & MIDAS Regressor &      $\mu$     &  $\alpha$   &  $ \beta$  &  $\theta$   &  $\omega_2$   &   $m$   &  LLF   &    BIC    \\ \hline
\noalign{\smallskip}
\multicolumn{9}{l}{\textit{Panel A: Model I and Model II}}
 \vspace{2mm}\\ 
 Brent oil       &  Fixed $RV$    &   4.684E-$04^{*}$ & $0.058^{***}$ & $0.921^{***}$ & $0.165^{***}$ & $5.478^{**}$ & $0.013^{***}$   & 11646        & $-23241$     \\
               &              &    (2.568E-04)    &  (0.005)      & (0.009)       & (0.014)       &   (2.416)    & (0.001)  & &  \\
               &  Rolling $RV^{(\rm{rw})}$  & 4.614E-$04^{*}$   & $0.059^{***}$ & $0.916^{***}$ & $0.173^{***}$ & $7.272^{**}$ & $0.011^{***} $   & 11647        & $-23243$          \\
               &              &    (2.565E-04)    &  (0.006)      & (0.011)       & (0.0123)      &   (3.172)    & (0.0015)  & & \\
\noalign{\smallskip} 
 WTI oil &  Fixed $RV$    & 5.056E-$04^{*}$   & $0.057^{***}$ & $0.931^{***}$ & $0.118^{***}$ &  $4.949$     & $0.019^{***}$   &  11006       & $-21961$      \\
               &              &    (2.802E-04)    &   (0.005)     &   (0.008)     &    (0.038)    &  (4.868)     &   (0.003) & & \\
               &  Rolling $RV^{(\rm{rw})}$  & 5.112E-$04^{*}$   & $0.058^{***}$ & $0.929^{***}$ & $0.129^{***}$ &  $5.744$     & $0.018^{***}$  &  11006       & $-21961$       \\
               &              &    (2.813E-04)    &   (0.005)     &   (0.009)     &    (0.034)    &  (5.427)     &   (0.003) & &  \\
\noalign{\smallskip}
\multicolumn{9}{l}{\textit{Panel B: Model III and Model IV (fixed-span version)}}
 \vspace{2mm}\\ 
 Brent oil        &  $GEPU$     &     4.008E-04       & $0.050^{***}$ & $0.948^{***}$ &   $0.062^{***}$    & $49.965^{*}$  &    $-1.491$E-05    & 11011    & $-21970$  \\
     &            &                    (2.502E-04)      &  (0.004)      & (0.004)       & (0.019)      &  (27.893)     &    (1.447E-04) & & \\
                  & $\Delta{GEPU}$ &    4.137E-04        & $0.048^{***}$ & $0.946^{***}$ & $0.006^{***}$ & $2.994^{**}$ & 4.526E-04    & 11612        & $-23174$  \\
               &          &    (2.552E-04)      &  (0.004)      & (0.005)       & (0.002)      &   (0.988)    & (7.419E-05)  & & \\
\noalign{\smallskip} 
 WTI oil     &  $GEPU$    &    4.832E-$04^{*}$  & $0.053^{***}$ & $0.942^{***}$ & $0.035^{**}$ & $49.912^{*}$   &  2.289E-04  & 11010          & $-21969$            \\
               &          &    (2.754E-04)      &  (0.005)      & (0.005)       & (0.015)      &  (45.697)      & (1.446E-04)  & &  \\
             & $\Delta{GEPU}$ &    4.521E-04        & $0.050^{***}$ & $0.941^{***}$ &$0.006^{***}$ & $3.011$     & 5.069E-$04^{***}$ & 10967   & $-21883$     \\
               &          &    (2.782E-04)      &  (0.005)      & (0.005)       & (0.002)      &   (1.202)      & (7.056E-05)  & &  \vspace{2mm} \\
\multicolumn{9}{l}{\textit{Panel C: Model V and Model VI (rolling-window version)}}
 \vspace{2mm}\\

 Brent oil          &  $GEPU^{\rm{(rw)}}$  &     4.008E-04       & $0.050^{***}$ & $0.948^{***}$ & $0.062^{***}$ & $49.965^{*}$ & $-1.491$E-05    & 11011    & $-21970$   \\
               &          &    (2.502E-04)      &  (0.004)      & (0.004)       & (0.019)       & (27.893)     & (1.447E-04) & & \\
             & $\Delta{GEPU}^{\rm{(rw)}}$ &    4.137E-04        & $0.048^{***}$ & $0.946^{***}$ & $0.006^{***}$ & $2.994^{**}$ & 4.526E-04  & 11612        & $-23174$     \\
               &          &    (2.552E-04)      &  (0.004)      & (0.005)       & (0.002)      &   (0.988)    & (7.419E-05)  & & \\
\noalign{\smallskip} 
 WTI oil          &  $GEPU^{\rm{(rw)}}$    &    4.750E-$04^{*}$  & $0.052^{***}$ & $0.944^{***}$ & $0.042^{***}$ & $49.952^{*}$   &  1.512E-04  & 11011    & $-21970$      \\
               &          &    (2.748E-04)      &  (0.004)      & (0.005)       & (0.016)      &  (32.444)      & (1.381E-04)  & &  \\
             & $\Delta{GEPU}^{\rm{(rw)}}$ &    4.521E-04        & $0.050^{***}$ & $0.942^{***}$ &$0.007^{***}$ & $3.148$      & 5.037E-$04^{***}$  & 10968    & $-21884$   \\
               &          &    (2.785E-04)      &  (0.005)      & (0.005)       & (0.002)      &   (1.105)      & (6.811E-05)  & & \\
 \hline 
\end{tabular}    
}   
\begin{flushleft}
\footnotesize     
\justifying Note: This table reports the parameter estimates of all the six single-factor models (Model I -- Model VI). The samples for Brent oil and WTI oil are both from 1 December 1998 to 31 October 2019. LLF is the log-likelihood function and BIC indicates the Bayesian information criterion. The numbers in the parentheses are the standard deviation. Superscripts $^{***}$, $^{**}$, and $^{*}$ denote respectively the significance levels at $1\%$, $5\%$, and $10\%$.
Panel A reports the parameter estimates of Model I and Model II. Fixed $RV$ denotes the realized volatility with fixed span, which is calculated from Eq.~(\ref{Eq:RV:fixspan}). Rolling-window $RV^{(\rm{rw})}$ represents the realized volatility with rolling window, which is calculated from Eq.~(\ref{Eq:RV:RW}). 
Panel B reports the parameter estimates of Model III and Model IV.
Panel C reports the parameter estimates of Model V and Model VI. The the GEPU index and its changes here use the rolling-window version, whose monthly value is copied to the value of each day of the corresponding month.
\end{flushleft}                                                  
\end{table}

\begin{figure}[htbp]
\centering
\includegraphics[width=0.49\linewidth]{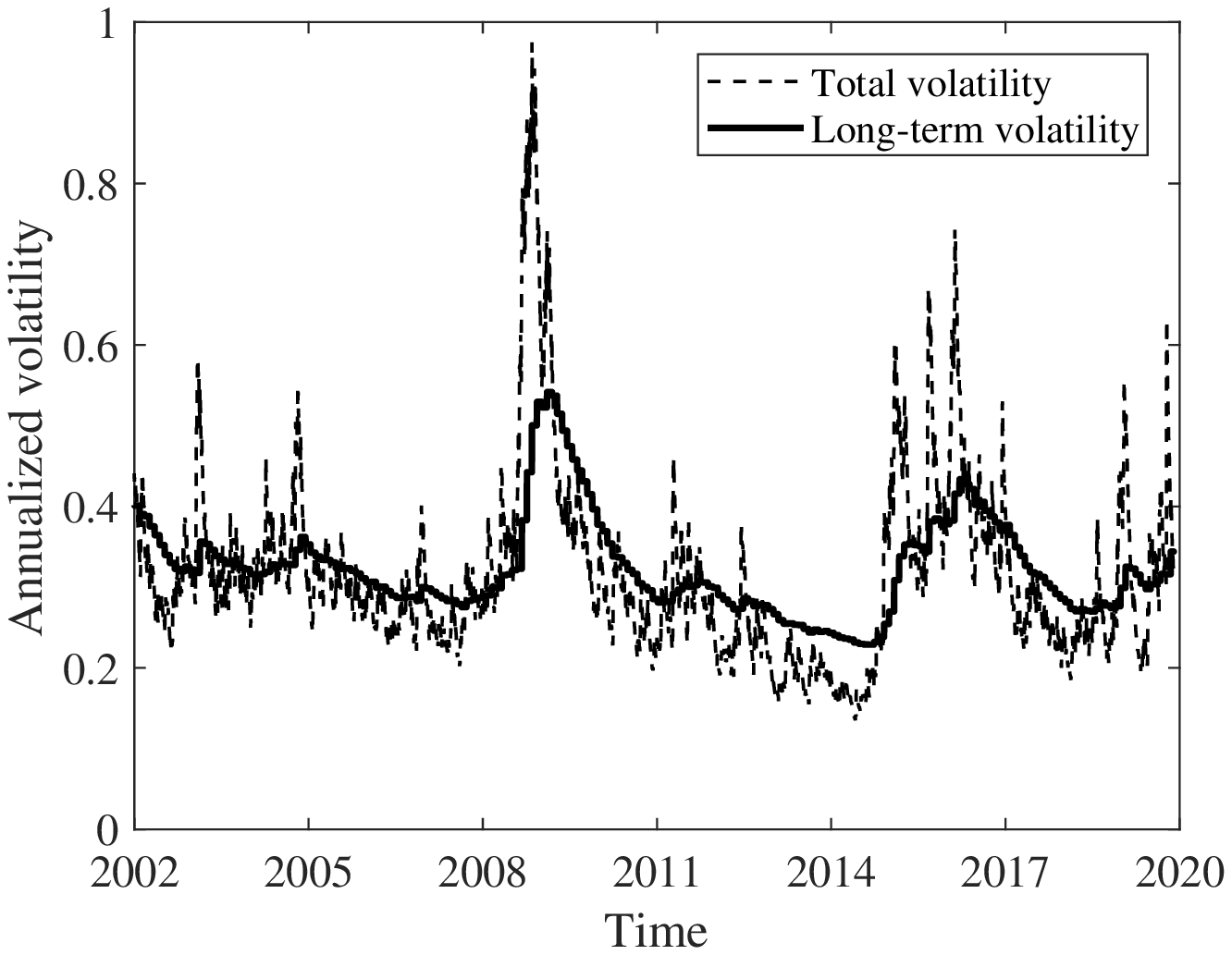}
\includegraphics[width=0.49\linewidth]{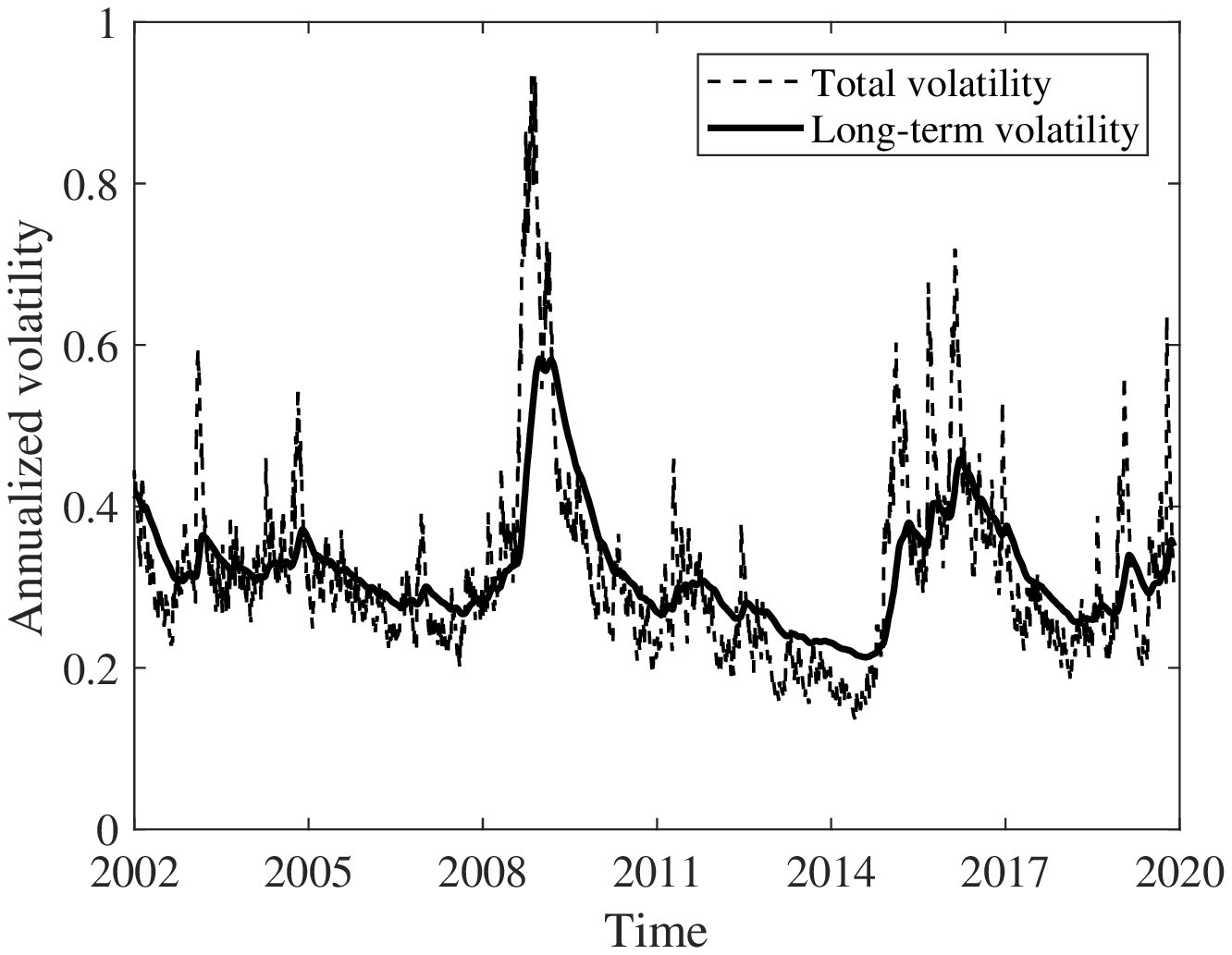}
\includegraphics[width=0.49\linewidth]{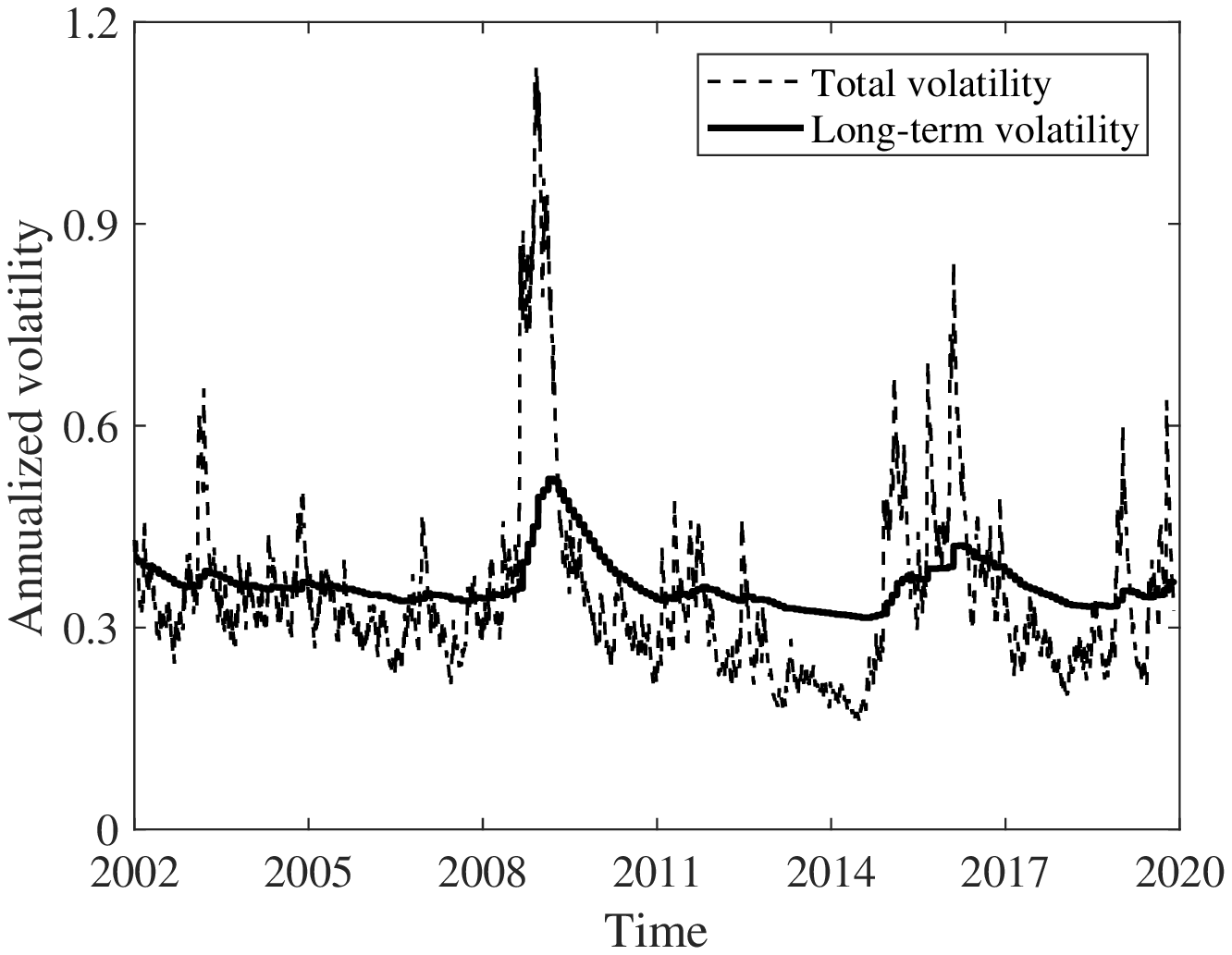}
\includegraphics[width=0.49\linewidth]{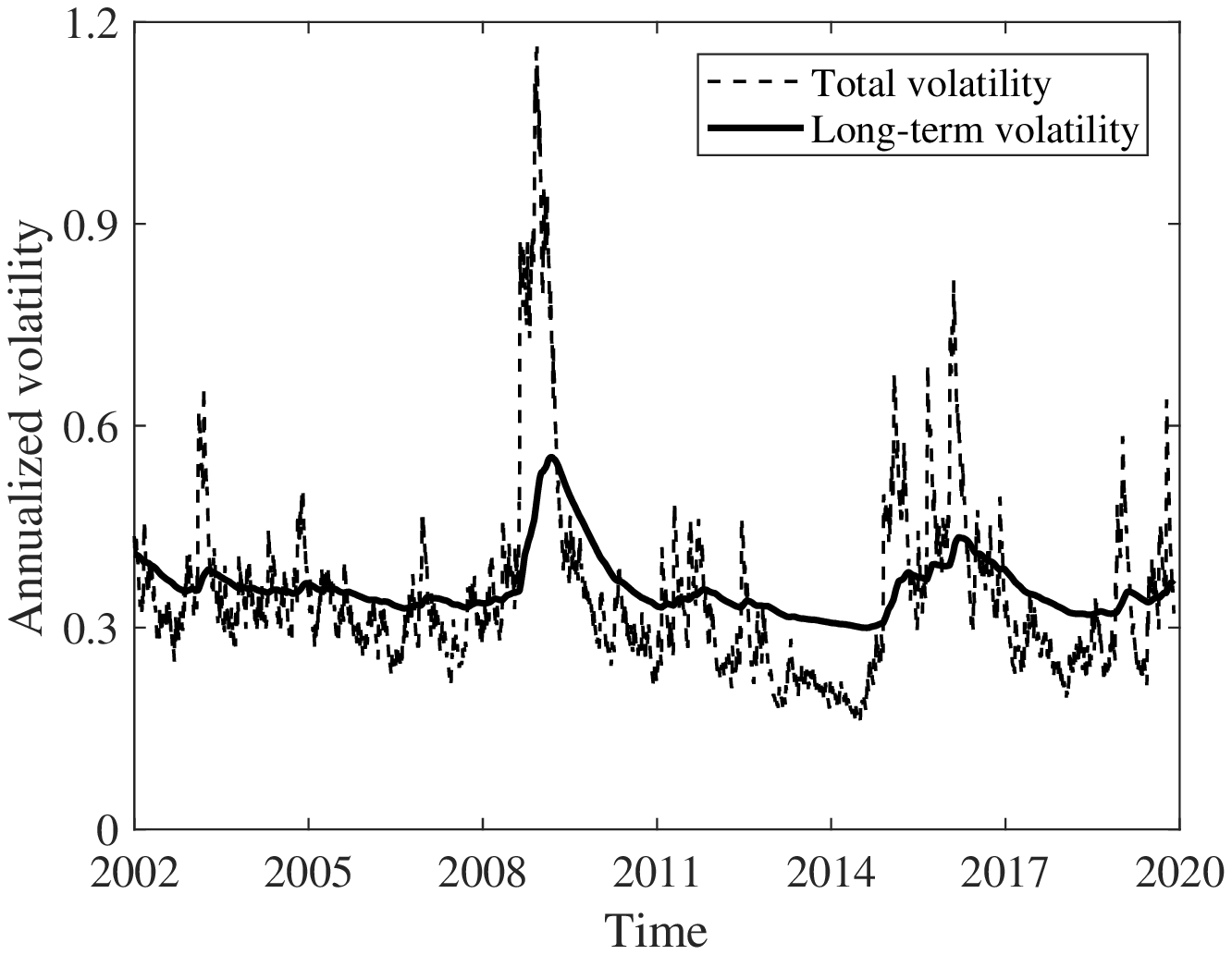}
\caption{Total volatility and its long-term component of crude oil futures return from single-factor model with realized volatility. The left column is realized volatility with fixed span (Model I) and the right column is realized volatility with rolling window (Model II). The top row is for Brent oil, while the bottom row is for WTI oil. The plots show standard deviations on an annualized scale.  Total volatility and its long-term component of crude oil futures return from single-factor models.}
\label{Fig:OilFutGEPU:SinFact:RV}
\end{figure}

Fig.~\ref{Fig:OilFutGEPU:SinFact:RV} illustrates the annualized long-term volatility and total volatility of the crude oil futures returns which are calculated from the single-factor models with fixed-span $RV$ and rolling-window $RV^{(\rm{rw})}$. As can be seen from the figure, the long-term volatility curve from Model II is smoother than that from Model I. The evolutionary trend of the long-term volatility is consistent with the corresponding total volatility. Certainly, the difference between them is also visible by eye-balling.

Next, we discuss the individual influence of economic policy uncertainty on crude oil futures volatility. The single-factor models (Model III to Model VI) satisfy our requirement to examine the effect of economic policy uncertainty. Similar to realized volatility, the GEPU index and its changes are set as two versions, fixed-span specification and rolling-window specification. The rolling-window specification about the macroeconomic variable (the GEPU index and its changes) is defined as follows:
\begin{equation}
  MV_i^{({\rm{rw}})}= \frac{1}{N'}\sum_{j=1}^{N'}MV_{i-j},
  \label{Eq:MV}
\end{equation}
where $MV_i$ is a daily variable and its value equals to the corresponding monthly value,  $MV_i^{({\rm{rw}})}$ is the mean of a month earlier before the $i$-th day. It's worth mentioning that we could carry out preciser analysis if we have the real daily data of economic policy uncertainty.

Model III and Model IV are single-factor models with fixed-span specification, while Model V and Model VI are single-factor models with rolling-window specification. The whole sample data is selected to estimate the parameters of the four models, which are also presented in Table ~\ref{TB:OilFutGEPU:Calibration:SinFactor}. Panel B of Table ~\ref{TB:OilFutGEPU:Calibration:SinFactor} reports the parameter estimates of single-factor models with fixed-span specification, while Panel C of Table~\ref{TB:OilFutGEPU:Calibration:SinFactor} reports the parameter estimates of single-factor models with rolling-window specification.

In Panel B and Panel C of Table~\ref{TB:OilFutGEPU:Calibration:SinFactor}, the values of $\alpha$ and $\beta$ are all significantly different from 0 at the $1\%$ level and all the $\beta>0.9$, which means the estimated short-term volatility from the single-factor model with economic policy uncertainty exhibits strong volatility clustering. Concerning the single-factor model with the GEPU index, the values of $\theta$ for the two commodity futures are significantly positive at the $5\%$ level both in the fixed-span version and rolling-window version, which implies that the long-term volatility of crude oil futures return responds to global uncertainty of economic policy positively. The long-term volatility is heavy when the economic policy uncertainty is high without considering the realized volatility. 
Similarly, the long-term volatility of both commodities futures respond to economic policy uncertainty changes towards the same direction. The GEPU changes have consistent contributions to the crude oil futures volatility since the $\theta$ values of the two commodities are significantly positive at the $1\%$ level both in the fixed-span version and rolling-window version. The long-term volatility of crude oil futures returns is heavy when the global economic policy uncertainty changes strongly without considering the realized volatility. Even the GEPU is at low level, obvious change will lead to heavy volatility. This is a quite interesting provisional result which implies our follow-up research meaningful. Comparing the results of the fixed-span version and the rolling-window version of each commodity in Table~\ref{TB:OilFutGEPU:Calibration:SinFactor}, we find that the BIC of the fixed-span version is not smaller than that of the rolling-window version. Therefore, the rolling-window version of the single-factor model is preciser than the fixed-span version, no matter the factor is the GEPU index or its changes.

\begin{figure}[htbp]
\centering
\includegraphics[width=0.49\linewidth]{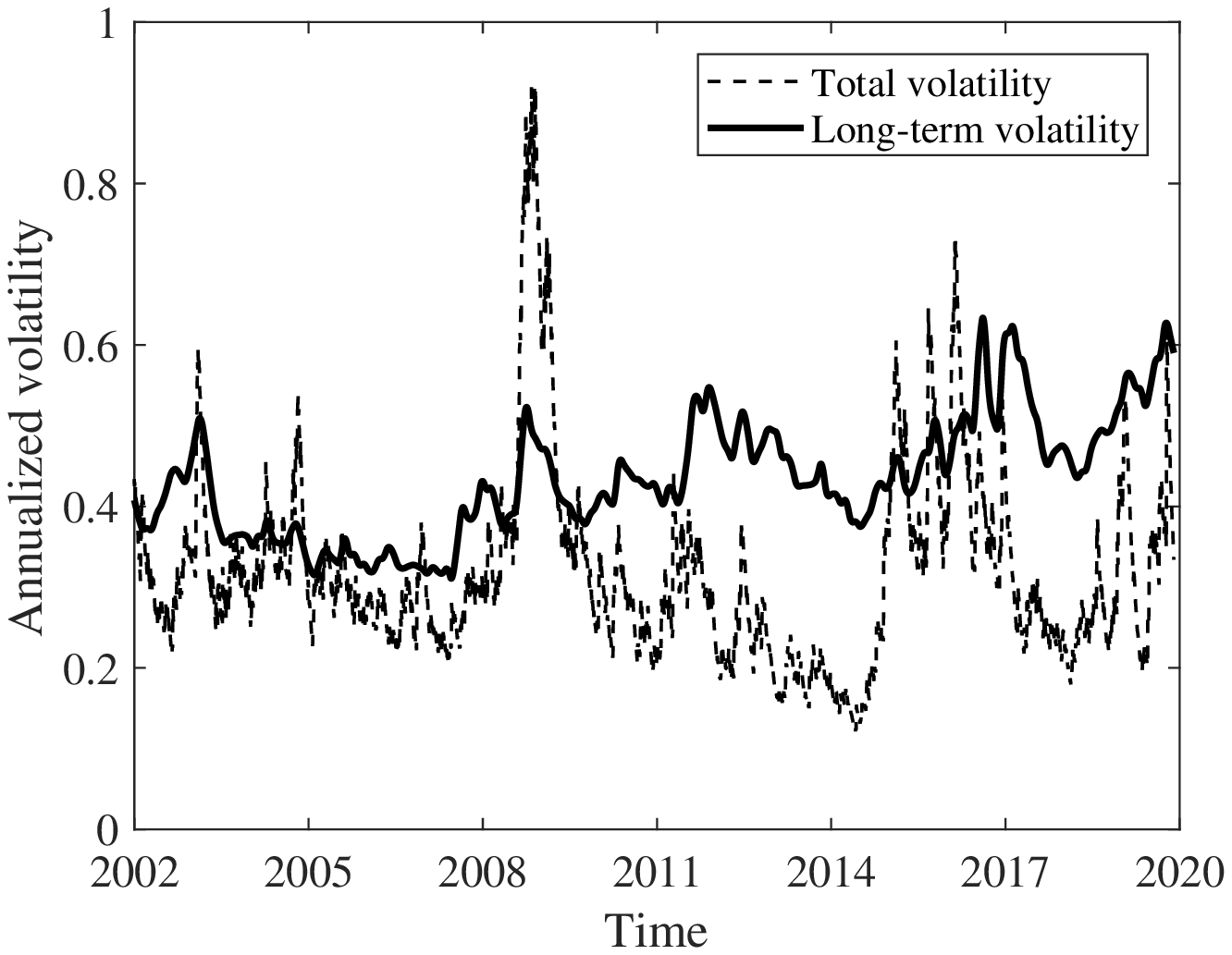}
\includegraphics[width=0.49\linewidth]{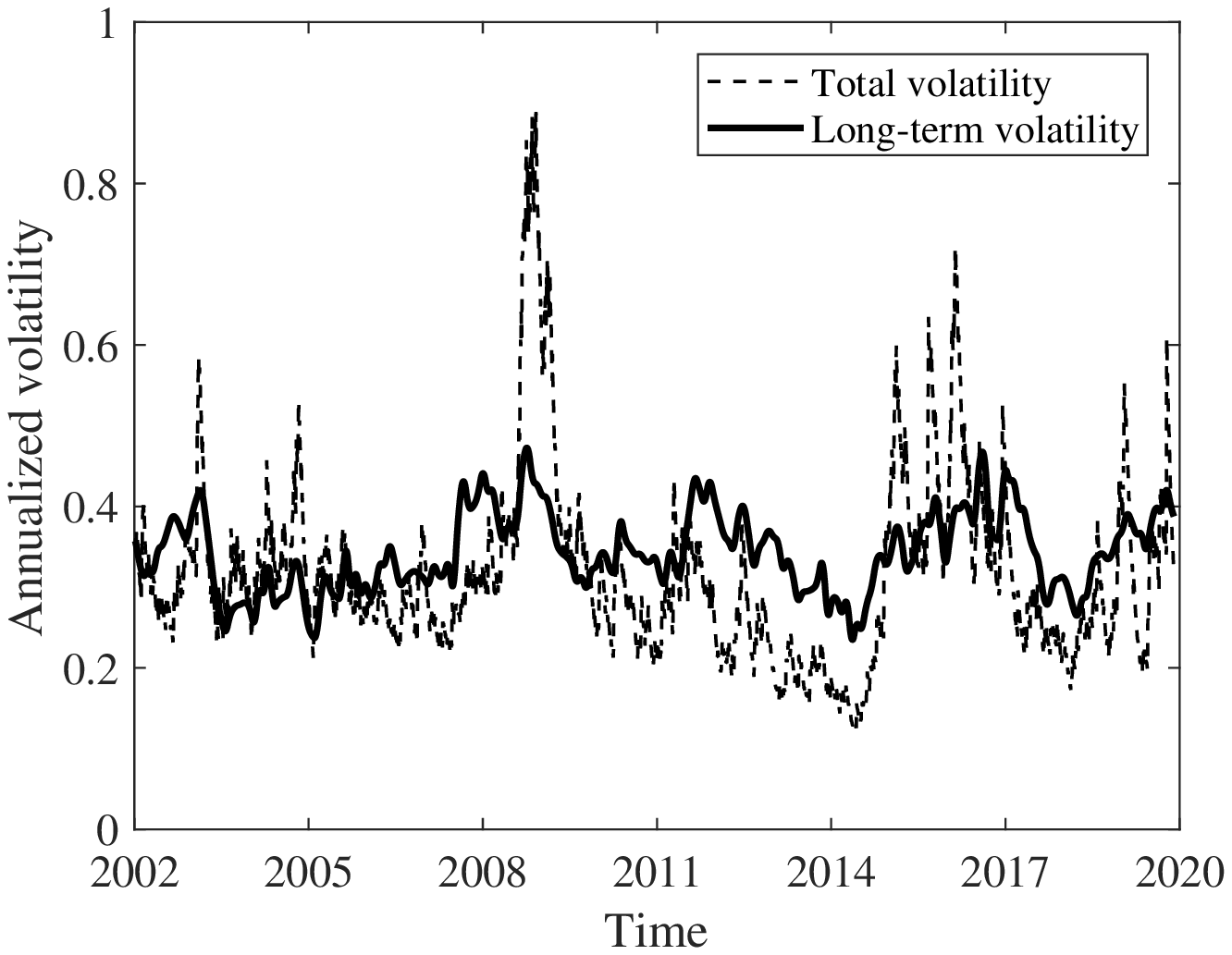}
\includegraphics[width=0.49\linewidth]{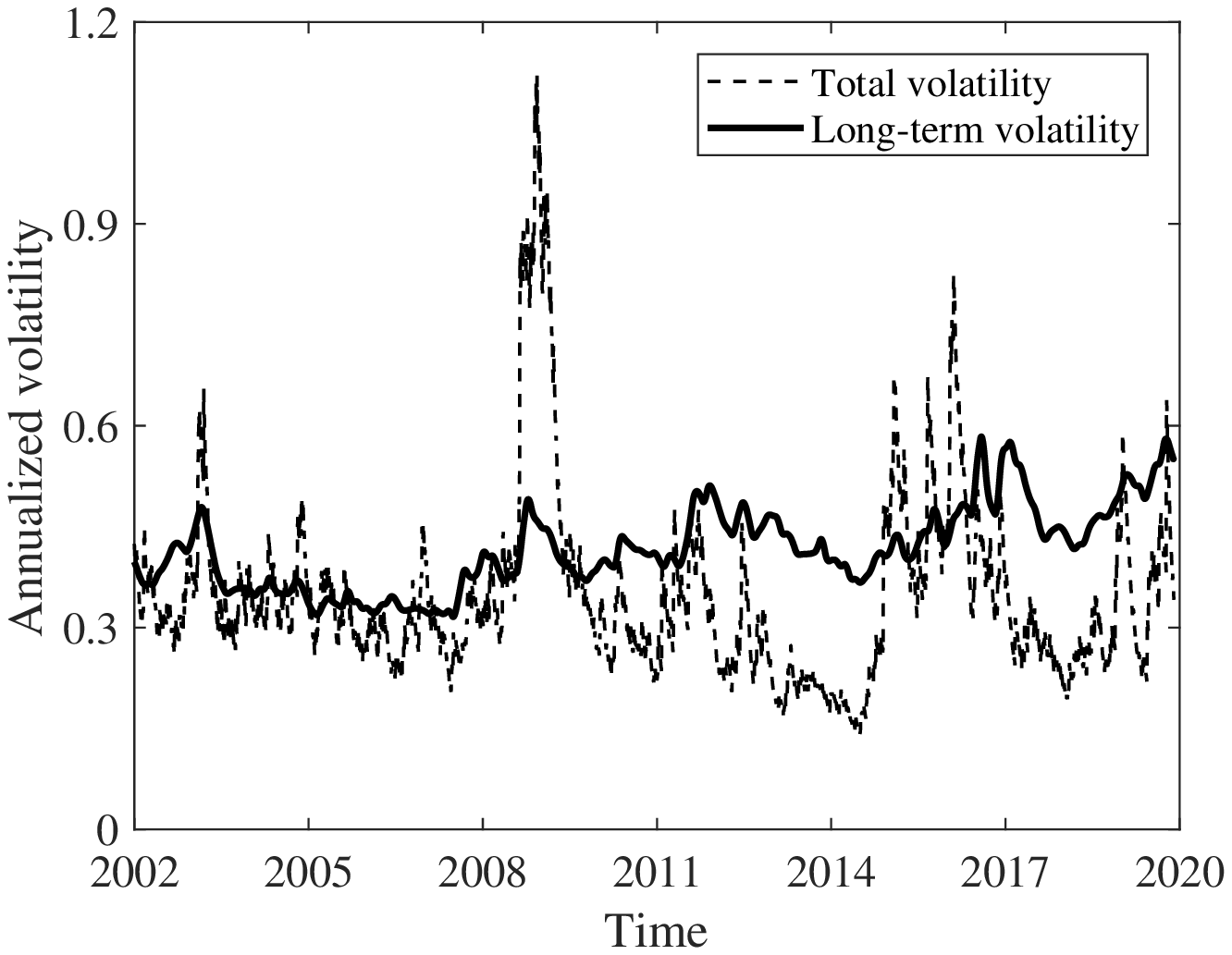}
\includegraphics[width=0.49\linewidth]{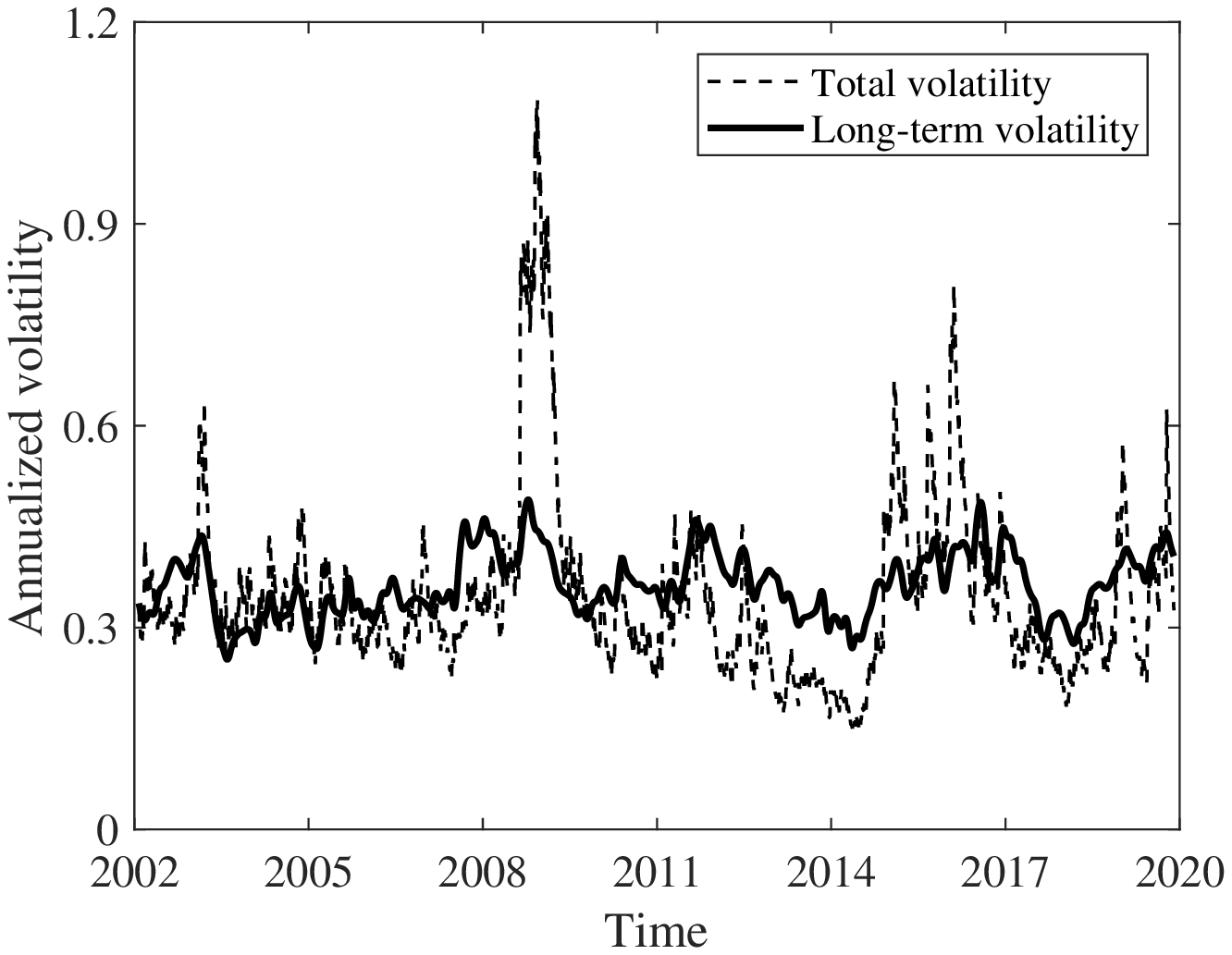}
\caption{Total volatility and its long-term component of crude oil futures estimated from Model V and Model VI. The left column is for the GEPU index (Model V) and the right column is for the GEPU changes (Model VI). The two rows display in turn the results for Brent oil and WTI oil. The plots show standard deviations on an annualized scale.}
\label{Fig:OilFutGEPU:SinFact:GEPU/GEPUC}
\end{figure}

Fig.~\ref{Fig:OilFutGEPU:SinFact:GEPU/GEPUC} illustrates the estimated annualized total volatility and its long-term component of crude oil futures derived from the rolling-window version of the single-factor model with the GEPU index and its changes (Model V and Model VI)\footnote{To save space, we do not present the results from the fixed-span version here.}.
The left column is for the GEPU index and the right column is for the GEPU changes. We find that the long-term volatility estimated from the single-factor model with GEPU changes (Model VI) is closer to its corresponding total volatility. In each plot, the pair of total volatility and long-term volatility evolve in a similar trend during the whole sample period. Thus, the short-term volatility is more sensitive to extreme events which usually cause drastic fluctuations of the crude oil futures market. Obvious separation also exists between the two trajectories of the total volatility and long-term volatility, induced by the short-term volatility. Extreme events also affect the development of long-term volatility, thus the long-term volatility follows cyclical pattern presented in the figure. 

The implication about the value of $\theta$ could be visually reviewed in Fig.~\ref{Fig:OilFutGEPU:SinFact:GEPU/GEPUC}. For example, the GEPU index and its changes rise around the global financial crisis of 2008, and the estimated contemporaneous long-term volatility from Model V and Model VI ascend simultaneously. By comparing the two columns, we discover that the long-term volatility related to the GEPU changes has higher resolutions than that related to the GEPU index. Looking back to the results of the single-factor models with realized volatility, we can see the long-term volatility related to the GEPU changes also has higher resolutions. Considering the period from 2011 to 2014, for instance, long-term volatility curve estimated from the single-factor models with realized volatility (Model I and Model II) is quite gentle compared with that from the single-factor models with GEPU changes (Model V and model VI) for both Brent oil and WTI oil. Hence, economic policy uncertainty changes act as a more effective factor in the single-factor model to explain the volatility of crude oil futures market.


\subsection{Calibration of two-factor models}
\label{ModEst:TwofactorMod}

Unlike the single-factor models, the two-factor models combine macroeconomic variable (GEPU index or GEPU changes) with realized volatility of commodity futures returns. We utilize the two-factor models in Section~\ref{S3:Method:TwofactorMod} to study the extra explanatory power of macroeconomic variable to long-term volatility of crude oil futures returns eliminating the influence of realized volatility. In line with the single-factor models, we select the full sample data to deduce the estimates of all parameters. In calibrating the two-factor models with GEPU changes, we eliminate the daily returns of the first month of the crude oil futures. The parameter estimates in the two-factor models with fixed span are listed in Panel A of Table ~\ref{TB:OilFutGEPU:Calibration:TwoFactor} and Panel B of Table~\ref{TB:OilFutGEPU:Calibration:TwoFactor} reports the parameter estimates in two-factor models with rolling window. As far as the rolling-window version of two-factor model is concerned, we calculate the realized volatility and macroeconomic variable with rolling-window specification.

\begin{table}[htp]
\centering
\setlength{\abovecaptionskip}{0pt}
\setlength{\belowcaptionskip}{10pt}
\caption{Parameter calibration of two-factor models}
\smallskip
\label{TB:OilFutGEPU:Calibration:TwoFactor}
\resizebox{\textwidth}{!}{
\begin{tabular}{llccccccccc}
  \hline \hline
 Commodity     &   MV     &      $\mu$       &  $\alpha$     &  $ \beta$     & $\theta_{RV}$ & $\theta_{MV}$ &  $\omega_2$ &   $m$    &   LLF       &  BIC  \\ \hline 
 \noalign{\smallskip}
 \multicolumn{9}{l}{\textit{Panel A: Model VII and Model VIII (fixed-span version)}}
 \vspace{2mm}\\                   
 Brent oil        &  $GEPU$     & 4.678E-$04^{*}$  & $0.058^{***}$ & $0.922^{***}$ & $0.027^{***}$ &  $-0.002$     &$5.038^{**}$ & 1.845E-$04^{***}$      
                                                                                                             & 11646       & $-23233$      \\
               &          & (2.570E-04)      &  (0.006)      & (0.009)       & (0.005)       &    (0.004)    & (2.230)     & (6.303E-05) & &\\
             & $\Delta{GEPU}$ & 4.234E-$04^{*}$  & $0.051^{***}$ & $0.929^{***}$ & $0.022^{***}$ & $0.003^{***}$ &$4.803^{***}$& 1.800E-$04^{***}$   
                                                                                                             &   11615     & $-23170$     \\
               &          &    (2.546E-04)   &  (0.005)      & (0.008)       & (0.005)       &   (0.001)     & (1.439)     & (3.759E-05)& & \\
\noalign{\smallskip} 
 WTI oil          &  $GEPU$      & 5.038E-$04^{*}$  & $0.056^{***}$ & $0.934^{***}$ & $0.014$       &  $-0.023^{*}$ &  $1.142$    & 6.588E-$04^{***}$ &  11008      & $-21957$      \\
               &          &  (2.789E-04)     &  (0.005)      & (0.006)       & (0.008)       &    (0.012)    & (0.808)     & (2.235E-04)& & \\
             & $\Delta{GEPU}$ & 4.571E-04        & $0.051^{***}$ & $0.936^{***}$ & $0.011$       & $0.005^{**}$  & $3.328^{**}$& 3.487E-$04^{***}$ & 10963    & $-21875$       \\
               &          &  (2.780E-04)     &  (0.005)      & (0.007)       & (0.008)       &   (0.002)     & (1.464)     & (9.119E-05) &&  \vspace{2mm}\\
 \multicolumn{9}{l}{\textit{Panel B: Model IX and Model X (rolling-window version)}}
 \vspace{2mm}\\
 Brent oil    &  $GEPU^{\rm{(rw)}}$    & 4.615E-$04^{*}$  & $0.059^{***}$ & $0.916^{***}$ & $0.030^{***}$ &  $-0.001$     &$7.235^{**}$ & 1.397E-$04^{***}$  & 11647   & $-23235$      \\
               &          & (2.566E-04)      &  (0.006)      & (0.012)       & (0.004)       &    (0.003)    & (3.147)     & (5.125E-05) & &\\
              
             & $\Delta{GEPU}^{\rm{(rw)}}$ & 4.133E-$04^{*}$  & $0.051^{***}$ & $0.930^{***}$ & $0.022^{***}$ & $0.003^{***}$ &$4.538^{***}$& 1.854E-$04^{***}$  &  11615    & $-23171$     \\
               &          &    (2.541E-04)   &  (0.005)      & (0.008)       & (0.005)       &   (0.001)     & (1.496)     & (4.227E-05)& & \\
\noalign{\smallskip} 

 WTI oil         &   $GEPU^{\rm{(rw)}}$     & 5.069E-$04^{*}$  & $0.057^{***}$ & $0.934^{***}$ & $0.013$       &  $-0.019^{*}$ &  $1.397$    & 6.262E-$04^{***}$  &  11008   & $-21955$  \\
               &          &  (2.795E-04)     &  (0.005)      & (0.006)       & (0.009)       &    (0.011)    & (1.154)     & (2.182E-04)& & \\
              
             & $\Delta{GEPU}^{\rm{(rw)}}$ & 4.633E-$04^{*}$  & $0.051^{***}$ & $0.933^{***}$ & $0.015^{**}$  & $0.005^{**}$  & $3.982^{**}$& 3.005E-$04^{***}$ & 10964    & $-21878$       \\
               &          &  (2.784E-04)     &  (0.005)      & (0.008)       & (0.007)       &   (0.002)     & (1.479)     & (7.442E-05) && \\  
\hline
\end{tabular}
}

\begin{flushleft}
\footnotesize     
\justifying Note: This table reports the parameter estimates of the four two-factor models (Model VII -- Model IX). The samples for Brent oil and WTI oil are both from 1 December 1998 to 31 October 2019. LLF is the log-likelihood function and BIC indicates Bayesian information criterion. The numbers in the parentheses are the standard deviation. The superscripts $^{***}$, $^{**}$, and $^{*}$ denote respectively the significance level at $1\%$, $5\%$, and $10\%$.
Panel A reports the parameter estimates of Model VII and Model VIII and Panel B reports the parameter estimates of Model IX and Model X. 
The GEPU index and its changes here use the rolling-window version, whose monthly value is copied to the value of each day of the corresponding month.
\end{flushleft} 
\end{table}

In Table~\ref{TB:OilFutGEPU:Calibration:TwoFactor}, the $\beta$ values for all the commodities are significantly positive at the $1\%$ level and their values are close to the corresponding estimates of the single-factor models in Table~\ref{TB:OilFutGEPU:Calibration:SinFactor}. The two-factor models describe the volatility clustering of crude oil futures returns' short-term volatility, which is consistent with the conclusion from traditional GARCH-class models \citep{Lv-Shan-2013-PA,Agnolucci-2009-EE,Ergen-Rizvanoghlu-2016-EE}. Comparing the last columns in Panel A and Panel B of Table~\ref{TB:OilFutGEPU:Calibration:TwoFactor}, we note that, for each commodity futures, BIC of two-factor model with fixed-span GEPU index is larger than that with rolling-window GEPU index for each commodity futures and BIC of two-factor model with fixed-span GEPU changes is also larger than that with rolling-window GEPU changes. The comparative results reveal that the two-factor models with rolling window is more competitive in explaining crude oil futures volatility. 


Now, we turn our attention to the most important parameter, the coefficient of the macroeconomic variables. The different coefficient signs of $GEPU$ or $GEPU^{\rm{(rw)}}$ mean that the GEPU index has an opposite effect on the volatility of crude oil futures when considering realized volatility. In Table~\ref{TB:OilFutGEPU:Calibration:TwoFactor}, the coefficients of $GEPU$ and $GEPU^{\rm{(rw)}}$, $\theta_{MV}$, is $-0.002$ and $-0.001$ for Brent oil. These values are not significantly different from 0. For WTI oil, the coefficients of $GEPU$ and $GEPU^{\rm{(rw)}}$ are $-0.023$ and $-0.019$ and they are significantly different from 0 at the $10\%$ level, but the corresponding $\theta_{RV}$ values in the same model are not significantly different from 0. The calibrated results of the two-factor models with the GEPU index show that Model VII and Model IX are not suitable for estimating the long-term volatility of both crude oil futures, Brent oil and WTI oil. 

Table~\ref{TB:OilFutGEPU:Calibration:TwoFactor} also shows that the coefficients $\theta_{MV}$ of the GEPU changes in the two-factor models for the two commodity futures are all significantly positive at least at the $5\%$ level. Meanwhile, the corresponding coefficients $\theta_{RV}$ of realized volatility in the same two-factor model with rolling window are significantly positive at least at the $5\%$ level as well (see Panel B of Table~\ref{TB:OilFutGEPU:Calibration:TwoFactor}). The results indicate that the two-factor model with GEPU changes works well for the crude oil futures market. In the two-factor model, the GEPU changes have a positive effect on the crude oil futures market volatility when the realized volatility is considered together. Moreover, we draw a conclusion that the two-factor model with GEPU changes performs better than the single-factor model with GEPU changes by comparing the values of BIC in Panel C of Table~\ref{TB:OilFutGEPU:Calibration:SinFactor} and in Panel B of Table~\ref{TB:OilFutGEPU:Calibration:TwoFactor}. 

\begin{figure}[htbp]
\centering
\includegraphics[width=0.49\linewidth]{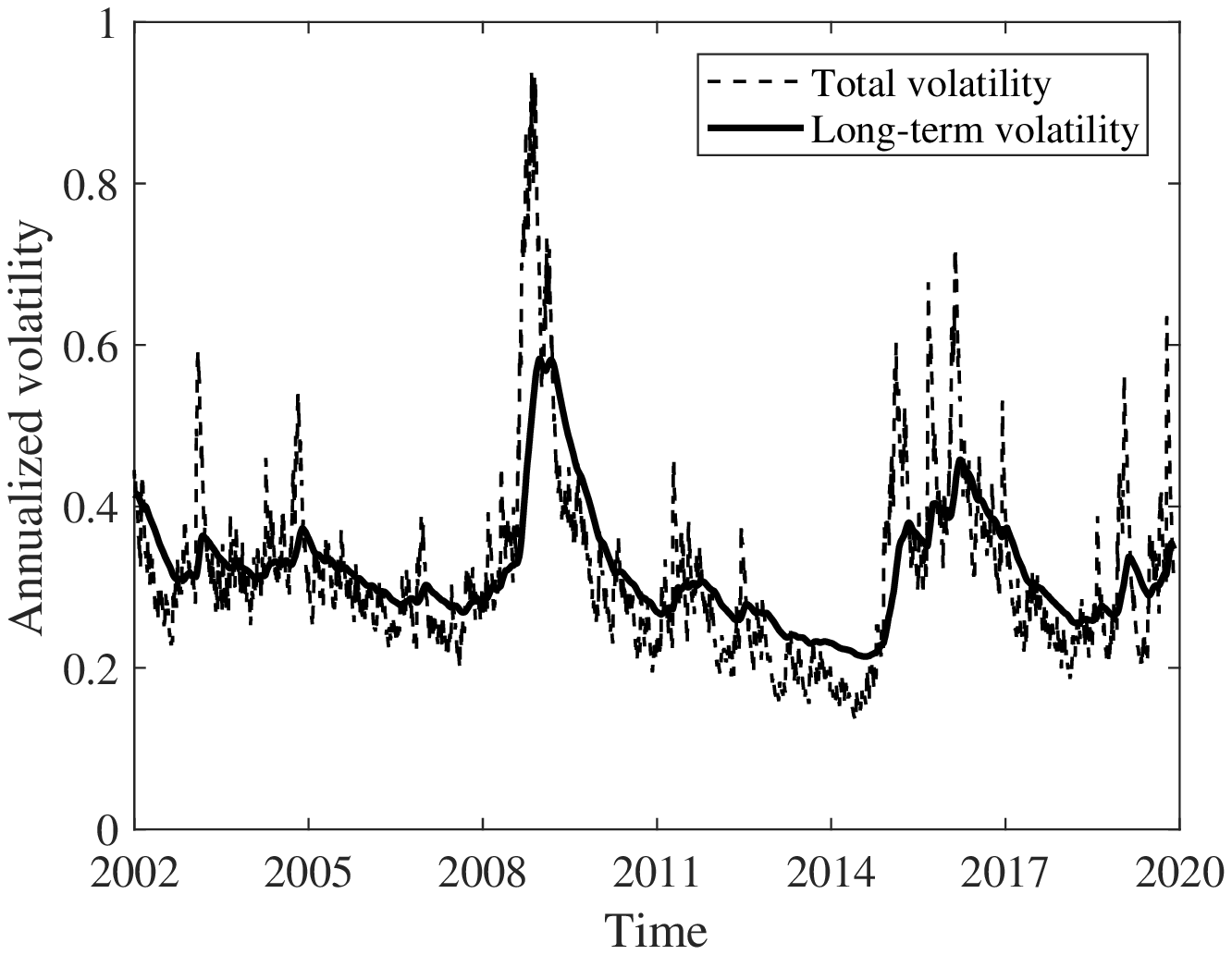}
\includegraphics[width=0.49\linewidth]{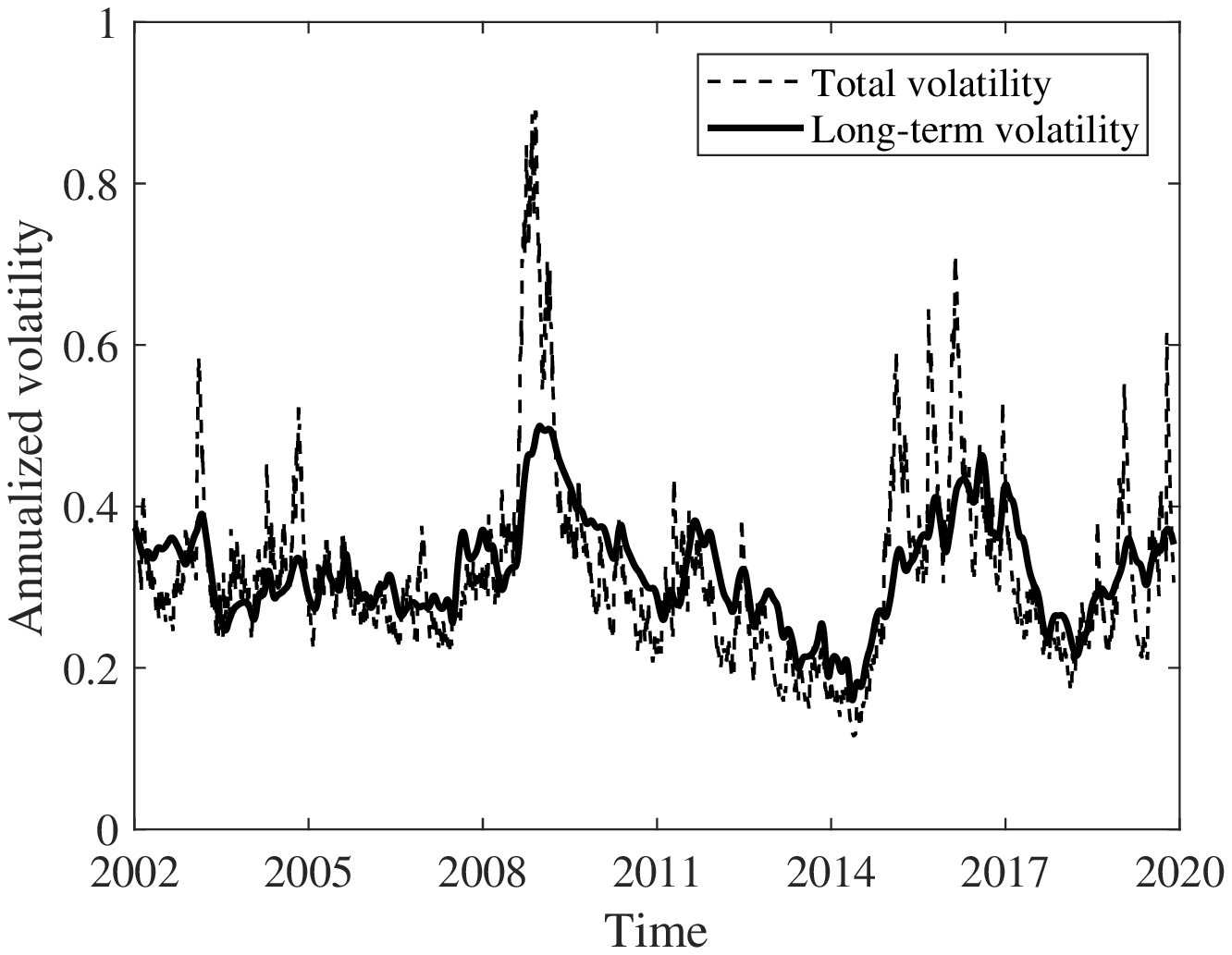}
\includegraphics[width=0.49\linewidth]{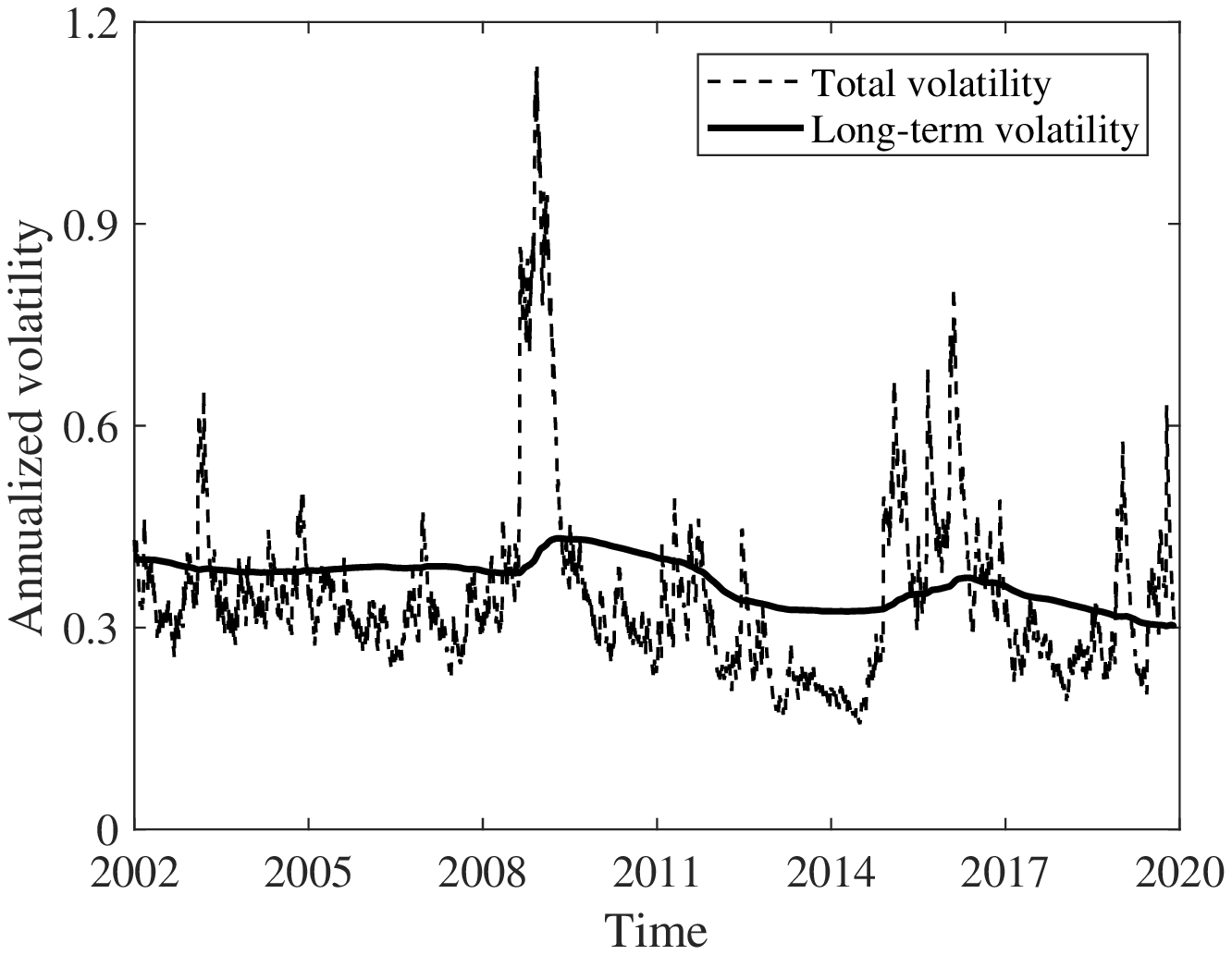}
\includegraphics[width=0.49\linewidth]{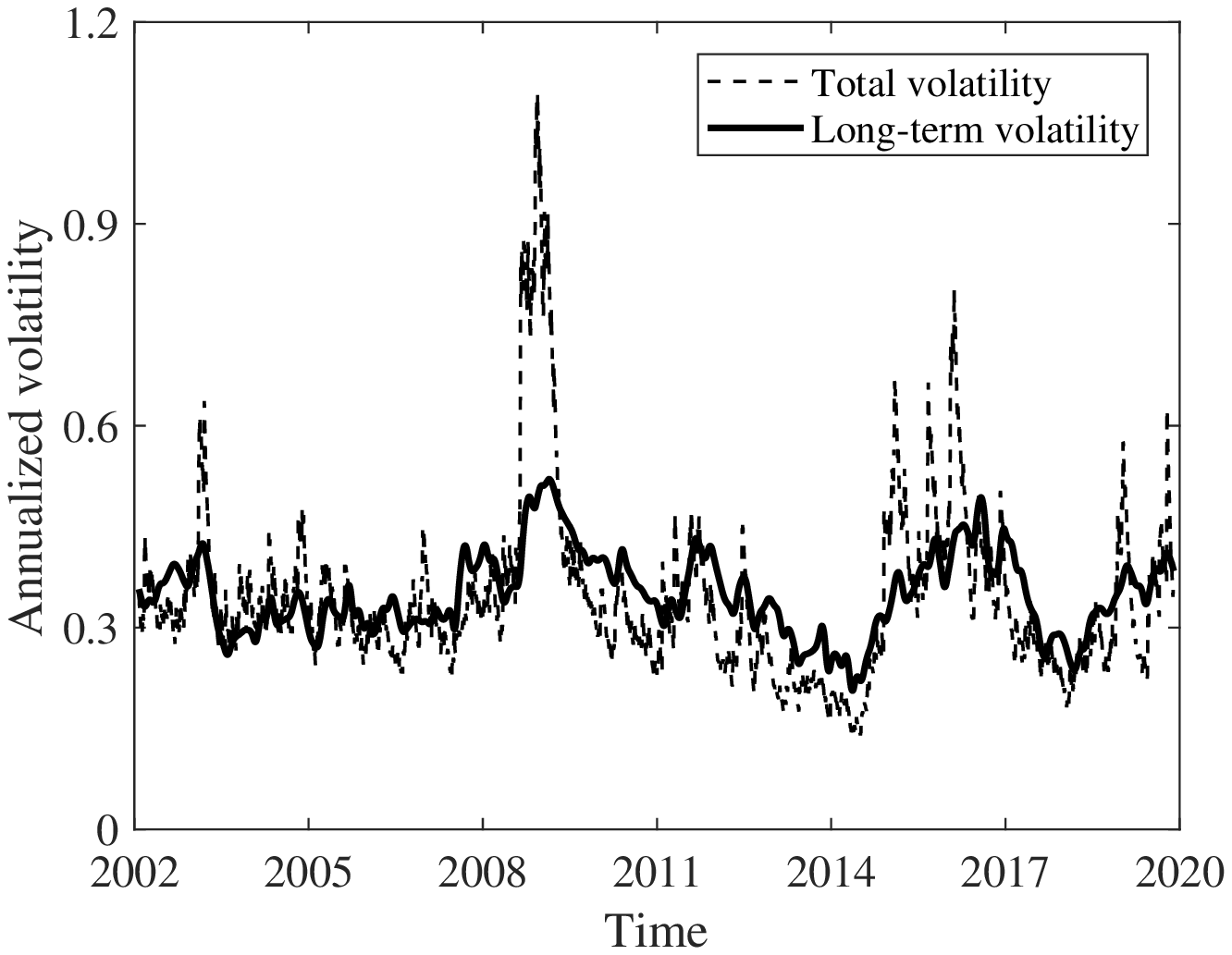}
\caption{Total volatility and its long-term component of crude oil futures estimated from Model IX and Model X. The left column is for the GEPU index and the right column is for the GEPU changes. The two rows display in turn the results for Brent oil and WTI oil. The volatility in each plot is annualized.}
\label{Fig:OilFutGEPU:TwoFact:RV+GEPU/GEPUS}
\end{figure}

In order to interpret the estimated results of the two-factor models intuitively, we plot the curves of total volatility and its long-term component of the two crude oil futures in Fig.~\ref{Fig:OilFutGEPU:TwoFact:RV+GEPU/GEPUS}. Visual inspection of the figure reveals that the two-factor model with GEPU changes provides better fits. Consistent with previous analyses, the estimated long-term volatility curve of WTI oil from the two-factor model with GEPU index does not characterize its real evolution trend precisely. For Brent oil, the estimated long-term volatility curve from the two-factor model with GEPU index between 2011 and 2015 is quite gentle while the curve from the two-factor model with GEPU changes has high resolutions. Focusing on the right columns in Fig.~\ref{Fig:OilFutGEPU:SinFact:RV}, Fig.~\ref{Fig:OilFutGEPU:SinFact:GEPU/GEPUC} and Fig.~\ref{Fig:OilFutGEPU:TwoFact:RV+GEPU/GEPUS}, we observe that the long-term volatility curves in Fig.~\ref{Fig:OilFutGEPU:TwoFact:RV+GEPU/GEPUS} is the closest to the corresponding total volatility curve and they have higher resolutions. Therefore, the two-factor model with GEPU changes is more appropriate to describe the crude oil futures volatility. 

\subsection{Evaluation results}
\label{ModEva}

In this section, we appraise the forecasting ability of all the models. To evaluate the variance prediction loss of a specific model, we select two popular loss functions mentioned in Section~\ref{S3:Method:ModEst&Eva}, $RMSE$ and $RMAE$, as the relevant measure. As the rolling-window version of the models preforms better than the corresponding fixed-span version, we just provide the evaluation results of the rolling-window version. The sample intervals of Brent oil and WTI oil are the same which covers from 1 December 1998 to 31 October 2019. 

In order to ensure a five-year sample period for out-of-sample prediction evaluation, we take a thirteen-year calibration window for WTI oil and Brent oil. Before the calibration window, there are additional three-year-lagged data needed to calculate the historical realized volatility. That is, we choose the data from period between 1 December 1998 and 31 December 2014 for the in-sample calibration and the data from period between 1 January 2015 and 31 October 2019 for the out-of-sample prediction. Table~\ref{TB:OilFutGEPU:Calibration:ModEva} not only reports the results of in-sample calibration and out-of-sample evaluation for WTI oil and Brent oil, but also lists the results of the full sample estimation in this table. 

\begin{table}[htp]
\centering
\setlength{\abovecaptionskip}{0pt}
\setlength{\belowcaptionskip}{10pt}
\caption{Results of model evaluation}
\smallskip
\label{TB:OilFutGEPU:Calibration:ModEva}
\resizebox{\textwidth}{!}{
\begin{tabular}{lccccccccc}
 \hline \hline
 \multirow{2}{*}{Commodity} & \multirow{2}{*}{Model}  &  \multicolumn{2}{c}{Full Sample} && \multicolumn{2}{c}{In-Sample}  & & \multicolumn{2}{c}{Out-of-Sample} \\   
 \cline{3-4} \cline{6-7} \cline{9-10}
    &   &   $RMSE$   &   $RMAE$     &&    $RMSE$     &     $RMAE$      &&     $RMSE$      &      $RMAE$  \\ \hline   
 \noalign{\smallskip}
 \multicolumn{9}{l}{\textit{Panel A: Single-factor models}} \\
\noalign{\smallskip}                                                                                    
 Brent oil   &      II    &   1.003E-03    &  2.198E-02       &&  9.772E-04  &   2.149E-02   &&   1.087E-03   &  2.346E-02    \\
    &   V   &   1.000E-03    &  2.249E-02       &&  9.733E-04  &   2.226E-02   &&   1.096E-03   &  2.394E-02      \\
    &  VI  &   9.953E-04    &  2.195E-02       &&  9.876E-04  &   2.095E-02   &&   1.089E-03   &  2.270E-02      \\ 
\noalign{\smallskip} 
 WTI oil        &  II       &   1.295E-03    &   2.395E-02      &&  1.323E-03  &   2.397E-02   &&   1.091E-03   &  2.458E-02     \\
             &   V    &   1.292E-03    &   2.421E-02      &&  1.323E-03  &   2.431E-02   &&   1.196E-03   &  2.496E-02     \\
             &   VI  &   1.282E-03    &   2.383E-02      &&  1.309E-03  &   2.367E-02   &&   1.191E-03   &  2.464E-02     \\ 
 \noalign{\smallskip}
 \multicolumn{9}{l}{\textit{Panel B: Two-factor models}}\\
\noalign{\smallskip}  
Brent oil      &    IX       &   1.003E-03    &  2.198E-02       &&  9.779E-04  &   2.145E-02   &&   1.088E-03    & 2.319E-02      \\
             &    X       &   9.972E-04    &  2.187E-02       &&  1.083E-03  &   2.296E-02   &&   8.713E-04    & 2.040E-02      \\ 
\noalign{\smallskip} 
                                            
WTI oil         &     IX      &   1.294E-03    &   2.394E-02      &&  1.325E-03  &   2.375E-02   &&   1.192E-03    &  2.431E-02     \\
               &     X    &   1.283E-03    &   2.380E-02      &&  1.310E-03  &   2.361E-02   &&   1.187E-03    &  2.449E-02     \\ 
\noalign{\smallskip}    
\hline                                         
\end{tabular}
}
\begin{flushleft}
\footnotesize     
\justifying Note: This table reports the results of full sample estimation, in-sample calibration and out-of-sample evaluation of the rolling-window models for WTI oil and Brent oil. The full samples of Brent oil and WTI oil are both from December 1998 to October 2019 and the out-of-sample prediction cover the period from November 2014 to October 2019.
Panel A reports the calculation results of loss functions ($RMSE$ and $RMAE$) for the single-factor models with rolling window (Model II, Model V and Model VI). 
Panel B reports the calculation results of loss functions ($RMSE$ and $RMAE$) for the two-factor models with rolling window (Model IX and Model X). 
\end{flushleft} 
\end{table}

The loss function values of the single-factor models with rolling window are presented in Panel A of Table~\ref{TB:OilFutGEPU:Calibration:ModEva}. Comparing the values of $RMSE$ and $RMAE$ between every two single-factor models, we find that Model VI causes less loss than other two models (Model II and Model V) when the full sample of Brent oil and WTI oil is concerned. The facts reveal that the GEPU changes is the more competitive factor to improve the interpretation capability of the single-factor model, compared with GEPU index and realized volatility. As far as the predictive ability of the single-factor model is concerned, the GEPU changes do not outperform other two factors, since the out-of-sample-values of $RMSE$ and $RMAE$ of Model VI are not always the smallest among the three models for the two commodity futures. However, the GEPU changes are indeed the volatility forecasting factor with better performance than the GEPU index. Panel B of Table~\ref{TB:OilFutGEPU:Calibration:ModEva} reports the loss function values of the two-factor models with rolling window. We do not find clear and unified numerical magnitude relationship between the out-of-sample-values of $RMAE$ of Model IX and Model X for the two commodity futures. 
In contrast, the full sample values and out-of-sample-values of $RMSE$ of Model X is smaller than that of Model IX for both commodity futures. This is strong evidence that the two-factor model with GEPU changes (Model X) is more effective to predict the crude oil futures volatility than the two-factor model with GEPU index (Model IX). Analysing the out-of-sample prediction error in Table~\ref{TB:OilFutGEPU:Calibration:ModEva}, we suggest that the two-factor model with GEPU changes contains more information and has stronger prediction power than the single-factor models.

 
The results of robustness test for the model evaluation are given in Table~\ref{TB:OilFutGEPU:TwoFactor:ModEvaTest}. As we can see from this table, the values of $RMSD$ and $RMAD$ of Model X are both smaller than that of Model IX, which implies that the two-factor model with GEPU changes is exactly more suitable to predict the crude oil futures volatility.

\begin{table}[htp]
\centering
\setlength{\abovecaptionskip}{0pt}
\setlength{\belowcaptionskip}{10pt}
\caption{Robustness test of two-factor model evaluation}
\smallskip
\label{TB:OilFutGEPU:TwoFactor:ModEvaTest}
\resizebox{\textwidth}{!}{
\begin{tabular}{cccccccccc}
 \hline \hline
 \multirow{2}{*}{Commodity} &   \multirow{2}{*}{Model}  &  \multicolumn{2}{c}{Full Sample} && \multicolumn{2}{c}{In-Sample}  & & \multicolumn{2}{c}{Out-of-Sample} \\   \cline{3-4} \cline{6-7} \cline{9-10}
           &                   &       $RMSD$     &     $RMAD$         &&    $RMSD$     &     $RMAD$      &&     $RMSD$      &      $RMAD$     \\ \hline                                       
\noalign{\smallskip}  
           
Brent oil   &   IX & 1.469E-02 &  0.1074 &&  1.524E-02 &  0.1095  && 1.403E-02 & 0.1057  \\
           &   X &   1.463E-02  & 0.1070 &&  1.521E-02  &  0.1092 && 1.392E-02  & 0.1051\\
\noalign{\smallskip} 
                                            
WTI oil     &  IX  & 1.590E-02 & 0.1110  &&  1.574E-02  &  0.1101  && 1.669E-02 &  0.1151   \\
            &   X  & 1.588E-02 & 0.1109  &&  1.570E-02  &  0.1098  && 1.668E-02 &  0.1150    \\
\hline                                                                                      
\end{tabular}
}
\begin{flushleft}
\footnotesize     
\justifying Note: This table shows the calculation results of loss functions ($RMSD$ and $RMAD$) for the two-factor models with rolling window (Model IX and Model X). The full samples of Brent oil and WTI oil are both from December 1998 to October 2019 and the out-of-sample prediction cover the period from November 2014 to October 2019.
\end{flushleft} 
\end{table}

Table~\ref{TB:OilFutGEPU:ModEva:DMTest} displays the results of DM test between every two models among all the models in the rolling-window version. It is worth noting that we do not compare the same models and the models with a common factor. We use ``$\times$" to represent these situation in Table~\ref{TB:OilFutGEPU:ModEva:DMTest}. According to this table, in term of the single-factor model, the predictive effect of the GEPU changes obviously outperforms that of the GEPU index since the corresponding $t$-statistic is $-5.85$ for Brent oil and $-3.42$ for WTI oil. Nevertheless, the realized volatility in the rolling-window version is the best predictive factor when considering the single-factor model. The forecasting capacity of the models is improved significantly (at the 1\% level) after adding the realized volatility to the single-factor model with GEPU index or its changes. On the other hand, the predictive power of the single-factor model with realized volatility is enhanced when integrating the GEPU index or its changes in the model. Summarizing the results in Table~\ref{TB:OilFutGEPU:ModEva:DMTest}, we can draw the conclusion that Model X has the best performance in predicting crude oil futures volatility.

\begin{table}[htp]
\centering
\setlength{\abovecaptionskip}{0pt}
\setlength{\belowcaptionskip}{10pt}
\caption{DM test for the out-of-sample performance of different models in rolling-window version}
\smallskip
\label{TB:OilFutGEPU:ModEva:DMTest}
\resizebox{\textwidth}{!}{
\begin{tabular}{lccccccccccc}
\hline \hline
      &  \multicolumn{5}{c}{Brent oil} && \multicolumn{5}{c}{WTI oil}  \\   \cline{2-6} \cline{8-12}
Model &    II        &    V       &    VI    &    IX    &   X     &   &     II   &    V    &    VI    &    IX    &    X      \\ \hline                                       
\noalign{\smallskip} 
 II   &  $\times$ &     $-$   &   $-$    &   $+$    & $+^{**}$  &&  $\times$  &   $-$     &   $-$    &   $+$    & $+^{**}$  \\
      &  $\times$ &  (-0.54)  & (-0.58)  & (1.28)   &  (1.98)   &&  $\times$  &  (-0.27)  & (-0.87)  & (1.43)   &  (2.02)   \\
 V    &   $+$     & $\times$  & $+^{***}$& $+^{***}$& $\times$  &&   $+$      & $\times$  & $+^{***}$& $+^{***}$& $\times$\\
      & (0.54)    &  $\times$ & (5.85)   & (3.15)   & $\times$  &&  (0.27)    & $\times$  &  (3.42)   & (2.80)   & $\times$   \\
 VI   &   $+$     &$-^{***}$  & $\times$ & $\times$ & $+^{***}$ &&   $+$      &$-^{***}$  & $\times$ & $\times$ & $+^{***}$   \\
      &   (0.58)  & (-5.85)   & $\times$ & $\times$ &  (3.06)   &&  (0.8 7)   & (-3.42)   & $\times$ & $\times$ &  (3.63) \\         
 IX   &   $-$     &$-^{***}$  & $\times$ & $\times$ &   $+$     &&   $-$      &$-^{***}$  & $\times$ & $\times$ &   $+$    \\                                           
      &  (-1.28)  &  (-3.15)  & $\times$ & $\times$ &  (1.25)   &&   (-1.43)  &  (-2.80)  & $\times$ & $\times$ &  (0.70)   \\
 X    & $-^{**}$  & $\times$  &$-^{***}$ &  $-$     & $\times$  &&  $-^{**}$  & $\times$  &$-^{***}$ &  $-$     & $\times$ \\
      & (-1.98)   & $\times$  & (-3.06)  &(-1.25)   & $\times$  &&  (-2.02)   & $\times$  & (-3.63)  &(-0.70)   & $\times$\\
\hline                                                                                      
\end{tabular}
}

\begin{flushleft}
\footnotesize     
\justifying Note: This table shows the DM test results between every two models among all the two-factor models in rolling-window version (Model II, Model V, Model VI, Model IX and Model X). The out-of-sample of Brent oil and WTI oil for DM test are from 1 November 2014 to 31 October 2019. ``$\times$" represents that the model corresponding to the row are not compared with models corresponding to the column. ``$-$" denotes the model corresponding to the row having higher accuracy compared with the model corresponding to the column, ``$+$" denotes the model corresponding to the row that owns lower accuracy compared with the model corresponding to the column. The numbers in the parentheses are the $t$-statistic. Subscripts $^{***}$, $^{**}$, and $^{*}$ denote being significance at the $1\%$, $5\%$, and $10\%$ levels respectively. 
\end{flushleft}

\end{table}

\section{Conclusions}
\label{S5:Conclusion}

In this work, we establish two types of models under the GARCH-MIDAS framework: single-factor models and two-factor models. Firstly, we employ the single-factor models to investigate respectively the impact of realized volatility, GEPU index and its changes on the crude oil futures volatility. Our empirical results show that the three factors produce significantly positive impacts on both commodity futures. From the perspective of single-factor models, the GEPU index and its changes are indeed the determinants of the crude oil futures volatility. A comparison of the competing models show that the GEPU changes outperform the GEPU index in predicting crude oil futures volatility. In addition, empirical results manifest that the single-factor models with rolling-window specification perform better than that with fixed-span specification. 

Then, we utilize the two-factor models to estimate respectively the impacts of GEPU index and its changes on the crude oil futures volatility when eliminating the effect of realized volatility. The calibration results of the two-factor model with GEPU changes reveal that GEPU changes can be a volatility forecasting factor for the crude oil futures market even when we include the realized volatility as an existing predictive factor. Moreover, our empirical results suggest that the two-factor model with GEPU changes is more suitable to describe the crude oil futures volatility. 
The increase of GEPU changes will result in the increase of the long-term volatility of crude oil futures. In addition, similar to the single-factor model, the two-factor model with rolling-window specification exhibits better performance. The findings of model evaluation indicate that the two-factor model with GEPU changes contains more information and has stronger predictive power for the crude oil futures volatility than the single-factor model. 

Our study indicates that the changes in global uncertainty of economic policy have significantly positive impacts on the long-term volatility of crude oil futures. We advise financial practitioners and policy makers to follow the guidance, taking global uncertainty changes into account when they want to predict the volatility of crude oil futures. This will help improve investment strategies and policy makers' decisions and probably lower or prevent the systemic risk of commodity markets. 

\section*{Acknowledgments}

This work was supported by National Natural Science Foundation of China (Grants Nos. 71532009, U1811462 and 71790594), Fundamental Research Funds for the Central Universities, Tianjin Development Program for Innovation and Entrepreneurship, and Program of Shanghai Academic Research Leader.

\bibliography{Bibliography}

\begin{thebibliography}{50}
\expandafter\ifx\csname natexlab\endcsname\relax\def\natexlab#1{#1}\fi
\providecommand{\url}[1]{\texttt{#1}}
\providecommand{\href}[2]{#2}
\providecommand{\path}[1]{#1}
\providecommand{\DOIprefix}{doi:}
\providecommand{\ArXivprefix}{arXiv:}
\providecommand{\URLprefix}{URL: }
\providecommand{\Pubmedprefix}{pmid:}
\providecommand{\doi}[1]{\href{http://dx.doi.org/#1}{\path{#1}}}
\providecommand{\Pubmed}[1]{\href{pmid:#1}{\path{#1}}}
\providecommand{\bibinfo}[2]{#2}
\ifx\xfnm\relax \def\xfnm[#1]{\unskip,\space#1}\fi
\bibitem[{Agnolucci(2009)}]{Agnolucci-2009-EE}
\bibinfo{author}{Agnolucci, P.}, \bibinfo{year}{2009}.
\newblock \bibinfo{title}{{Volatility in crude oil futures: A comparison of the
  predictive ability of GARCH and implied volatility models}}.
\newblock \bibinfo{journal}{Energy Econ.} \bibinfo{volume}{31},
  \bibinfo{pages}{316--321}.
\newblock \DOIprefix\doi{10.1016/j.eneco.2008.11.001}.
\bibitem[{Ames et~al.(2020)Ames, Bagnarosa, Matsui, Peters and
  Shevchenko}]{Ames-Bagnarosa-Matsui-Peters-Shevchenko-2020-EE}
\bibinfo{author}{Ames, M.}, \bibinfo{author}{Bagnarosa, G.},
  \bibinfo{author}{Matsui, T.}, \bibinfo{author}{Peters, G.W.},
  \bibinfo{author}{Shevchenko, P.V.}, \bibinfo{year}{2020}.
\newblock \bibinfo{title}{{Which risk factors drive oil futures price curves?}}
\newblock \bibinfo{journal}{Energy Econ.} \bibinfo{volume}{87},
  \bibinfo{pages}{104676}.
\newblock \DOIprefix\doi{10.1016/j.eneco.2020.104676}.
\bibitem[{Arbatli et~al.(2017)Arbatli, Davis, Ito, Miake and
  Saito}]{Arbatli-Davis-Ito-Miake-Saito-2017-IMF}
\bibinfo{author}{Arbatli, E.C.}, \bibinfo{author}{Davis, S.J.},
  \bibinfo{author}{Ito, A.}, \bibinfo{author}{Miake, N.},
  \bibinfo{author}{Saito, I.}, \bibinfo{year}{2017}.
\newblock \bibinfo{title}{{Policy uncertainty in Japan}}.
\newblock \bibinfo{journal}{IMF Working Paper}
  \DOIprefix\doi{10.5089/9781484300671.001}.
\bibitem[{Arouri et~al.(2011)Arouri, Jouini and
  Nguyen}]{Arouri-Jouini-Nguyen-2011-JIMF}
\bibinfo{author}{Arouri, M.E.H.}, \bibinfo{author}{Jouini, J.},
  \bibinfo{author}{Nguyen, D.K.}, \bibinfo{year}{2011}.
\newblock \bibinfo{title}{{Volatility spillovers between oil prices and stock
  sector returns: Implications for portfolio management}}.
\newblock \bibinfo{journal}{J. Int. Money Financ.} \bibinfo{volume}{30},
  \bibinfo{pages}{1387–1405}.
\newblock \DOIprefix\doi{10.1016/j.jimonfin.2011.07.008}.
\bibitem[{Asgharian et~al.(2013)Asgharian, Hou and
  Javed}]{Asgharian-Hou-Javed-2013-JFc}
\bibinfo{author}{Asgharian, H.}, \bibinfo{author}{Hou, A.},
  \bibinfo{author}{Javed, F.}, \bibinfo{year}{2013}.
\newblock \bibinfo{title}{{The importance of the macroeconomic variables in
  forecasting stock return variance: A GARCH-MIDAS approach}}.
\newblock \bibinfo{journal}{J. Forecast.} \bibinfo{volume}{32},
  \bibinfo{pages}{600--612}.
\newblock \DOIprefix\doi{10.1002/for.2256}.
\bibitem[{Bakas and Triantafyllou(2019)}]{Bakas-Triantafyllou-2019-EE}
\bibinfo{author}{Bakas, D.}, \bibinfo{author}{Triantafyllou, A.},
  \bibinfo{year}{2019}.
\newblock \bibinfo{title}{{Volatility forecasting in commodity markets using
  macro uncertainty}}.
\newblock \bibinfo{journal}{Energy Econ.} \bibinfo{volume}{81},
  \bibinfo{pages}{79--94}.
\newblock \DOIprefix\doi{10.1016/j.eneco.2019.03.016}.
\bibitem[{Baker et~al.(2013)Baker, Bloom and
  Davis}]{Baker-Bloom-Davis-2013-CBRP}
\bibinfo{author}{Baker, S.R.}, \bibinfo{author}{Bloom, N.},
  \bibinfo{author}{Davis, S.J.}, \bibinfo{year}{2013}.
\newblock \bibinfo{title}{{Measuring economic policy uncertainty}}.
\newblock \bibinfo{note}{{C}hicago Booth Research Paper}.
\bibitem[{Baker et~al.(2016)Baker, Bloom and
  Davis}]{Baker-Bloom-Davis-2016-QJE}
\bibinfo{author}{Baker, S.R.}, \bibinfo{author}{Bloom, N.},
  \bibinfo{author}{Davis, S.J.}, \bibinfo{year}{2016}.
\newblock \bibinfo{title}{{Measuring economic policy uncertainty}}.
\newblock \bibinfo{journal}{Quart. J. Econ.} \bibinfo{volume}{131},
  \bibinfo{pages}{1593--1636}.
\newblock \DOIprefix\doi{10.1093/qje/qjw024}.
\bibitem[{Bloom(2009)}]{Bloom-2009-Em}
\bibinfo{author}{Bloom, N.}, \bibinfo{year}{2009}.
\newblock \bibinfo{title}{{The impact of uncertainty shocks}}.
\newblock \bibinfo{journal}{Econometrica} \bibinfo{volume}{77},
  \bibinfo{pages}{623--685}.
\newblock \DOIprefix\doi{10.3982/ECTA6248}.
\bibitem[{Castelnuovo and Tran(2017)}]{Castelnuovo-Tran-2017-EL}
\bibinfo{author}{Castelnuovo, E.}, \bibinfo{author}{Tran, T.D.},
  \bibinfo{year}{2017}.
\newblock \bibinfo{title}{{Google it up! A google trends-based uncertainty
  index for the United States and Australia}}.
\newblock \bibinfo{journal}{Econ. Lett.} \bibinfo{volume}{161},
  \bibinfo{pages}{149--153}.
\newblock \DOIprefix\doi{10.1016/j.econlet.2017.09.032}.
\bibitem[{Chang and Lee(2015)}]{Chang-Lee-2015-EE}
\bibinfo{author}{Chang, C.P.}, \bibinfo{author}{Lee, C.C.},
  \bibinfo{year}{2015}.
\newblock \bibinfo{title}{{Do oil spot and futures prices move together?}}
\newblock \bibinfo{journal}{Energy Econ.} \bibinfo{volume}{50},
  \bibinfo{pages}{379--390}.
\newblock \DOIprefix\doi{10.1016/j.eneco.2015.02.014}.
\bibitem[{Cheng et~al.(2018)Cheng, Nikitopoulos and
  Schlogl}]{Cheng-Nikitopoulos-Schlogl-2018-JBF}
\bibinfo{author}{Cheng, B.}, \bibinfo{author}{Nikitopoulos, C.S.},
  \bibinfo{author}{Schlogl, E.}, \bibinfo{year}{2018}.
\newblock \bibinfo{title}{{Pricing of long-dated commodity derivatives: Do
  stochastic interest rates matter?}}
\newblock \bibinfo{journal}{J. Bank. Financ.} \bibinfo{volume}{95},
  \bibinfo{pages}{148--166}.
\newblock \DOIprefix\doi{10.1016/j.jbankfin.2017.05.012}.
\bibitem[{Dai et~al.(2019)Dai, Xiong and Zhou}]{Dai-Xiong-Zhou-2019-PA}
\bibinfo{author}{Dai, P.F.}, \bibinfo{author}{Xiong, X.},
  \bibinfo{author}{Zhou, W.X.}, \bibinfo{year}{2019}.
\newblock \bibinfo{title}{{Visibility graph analysis of economy policy
  uncertainty indices}}.
\newblock \bibinfo{journal}{Physica A} \bibinfo{volume}{531},
  \bibinfo{pages}{121748}.
\newblock \DOIprefix\doi{10.1016/j.physa.2019.121748}.
\bibitem[{Dai et~al.(2020)Dai, Xiong and Zhou}]{Dai-Xiong-Zhou-2020-FRL}
\bibinfo{author}{Dai, P.F.}, \bibinfo{author}{Xiong, X.},
  \bibinfo{author}{Zhou, W.X.}, \bibinfo{year}{2020}.
\newblock \bibinfo{title}{{A global economic policy uncertainty index from
  principal component analysis}}.
\newblock \bibinfo{journal}{Financ. Res. Lett.} ,
  \bibinfo{pages}{101686}\DOIprefix\doi{10.1016/j.frl.2020.101686}.
\bibitem[{Davis(2016)}]{Davis-2016-NBER}
\bibinfo{author}{Davis, S.}, \bibinfo{year}{2016}.
\newblock \bibinfo{title}{{An index of global economic policy uncertainty}}.
\newblock \bibinfo{type}{Technical Report}. NBER Working Paper No.22740.
\newblock \DOIprefix\doi{10.3386/w22740}.
\bibitem[{Diebold and Mariano(2002)}]{Diebold-Mariano-2002-JBES}
\bibinfo{author}{Diebold, F.X.}, \bibinfo{author}{Mariano, R.S.},
  \bibinfo{year}{2002}.
\newblock \bibinfo{title}{{Comparing predictive accuracy}}.
\newblock \bibinfo{journal}{J. Bus. Econ. Stat.} \bibinfo{volume}{20},
  \bibinfo{pages}{134--144}.
\newblock \DOIprefix\doi{10.1198/073500102753410444}.
\bibitem[{Engle and Rangel(2008)}]{Engle-Rangel-2008-RFS}
\bibinfo{author}{Engle, R.}, \bibinfo{author}{Rangel, J.G.},
  \bibinfo{year}{2008}.
\newblock \bibinfo{title}{{The spline-GARCH model for low-frequency volatility
  and its global macroeconomic causes}}.
\newblock \bibinfo{journal}{Rev. Financ. Stud.} \bibinfo{volume}{21},
  \bibinfo{pages}{1187--1222}.
\newblock \DOIprefix\doi{10.1093/rfs/hhn004}.
\bibitem[{Engle et~al.(2013)Engle, Ghysels and
  Sohn}]{Engle-Ghysels-Sohn-2013-RES}
\bibinfo{author}{Engle, R.F.}, \bibinfo{author}{Ghysels, E.},
  \bibinfo{author}{Sohn, B.}, \bibinfo{year}{2013}.
\newblock \bibinfo{title}{{Stock market volatility and macroeconomic
  fundamentals}}.
\newblock \bibinfo{journal}{Rev. Econ. Stat.} \bibinfo{volume}{95},
  \bibinfo{pages}{776--797}.
\newblock \DOIprefix\doi{10.1162/REST\_a\_00300}.
\bibitem[{Ergen and Rizvanoghlu(2016)}]{Ergen-Rizvanoghlu-2016-EE}
\bibinfo{author}{Ergen, I.}, \bibinfo{author}{Rizvanoghlu, I.},
  \bibinfo{year}{2016}.
\newblock \bibinfo{title}{{Asymmetric impacts of fundamentals on the natural
  gas futures volatility: An augmented GARCH approach}}.
\newblock \bibinfo{journal}{Energy Econ.} \bibinfo{volume}{56},
  \bibinfo{pages}{64--74}.
\newblock \DOIprefix\doi{10.1016/j.eneco.2016.02.022}.
\bibitem[{Fang et~al.(2019)Fang, Bouri, Gupta and
  Roubaud}]{Fang-Bouri-Gupta-Roubaud-2019-IRFA}
\bibinfo{author}{Fang, L.}, \bibinfo{author}{Bouri, E.},
  \bibinfo{author}{Gupta, G.}, \bibinfo{author}{Roubaud, D.},
  \bibinfo{year}{2019}.
\newblock \bibinfo{title}{{Does global economic uncertainty matter for the
  volatility and hedging effectiveness of bitcoin?}}
\newblock \bibinfo{journal}{Int. Rev. Financ. Anal.} \bibinfo{volume}{61},
  \bibinfo{pages}{29--36}.
\newblock \DOIprefix\doi{10.1016/j.irfa.2018.12.010}.
\bibitem[{Fang et~al.(2018)Fang, Chen, Yu and
  Qian}]{Fang-Chen-Yu-Qian-2018-JFutM}
\bibinfo{author}{Fang, L.}, \bibinfo{author}{Chen, B.}, \bibinfo{author}{Yu,
  H.}, \bibinfo{author}{Qian, Y.}, \bibinfo{year}{2018}.
\newblock \bibinfo{title}{{The importance of global economic policy uncertainty
  in predicting gold futures market volatility: A GARCH-MIDAS approach}}.
\newblock \bibinfo{journal}{J. Fut. Markets} \bibinfo{volume}{38},
  \bibinfo{pages}{413--422}.
\newblock \DOIprefix\doi{10.1002/fut.21897}.
\bibitem[{Fazelabdolabadi(2019)}]{Fazelabdolabadi-2019-FinancInnov}
\bibinfo{author}{Fazelabdolabadi, B.}, \bibinfo{year}{2019}.
\newblock \bibinfo{title}{Uncertainty and energy-sector equity returns in
  {I}ran: a {B}ayesian and quasi-{M}onte {C}arlo time-varying analysis}.
\newblock \bibinfo{journal}{Financ. Innov.} \bibinfo{volume}{5},
  \bibinfo{pages}{12}.
\newblock \DOIprefix\doi{10.1186/s40854-019-0128-2}.
\bibitem[{Geman and Kharoubi(2008)}]{Geman-Kharoubi-2008-JBF}
\bibinfo{author}{Geman, H.}, \bibinfo{author}{Kharoubi, C.},
  \bibinfo{year}{2008}.
\newblock \bibinfo{title}{{WTI crude oil futures in portfolio diversification:
  The time-to-maturity effect}}.
\newblock \bibinfo{journal}{J. Bank. Financ.} \bibinfo{volume}{32},
  \bibinfo{pages}{2553--2559}.
\newblock \DOIprefix\doi{10.1016/j.jbankfin.2008.04.002}.
\bibitem[{Ghysels et~al.(2007)Ghysels, Arthur and
  Rossen}]{Ghysels-Arthur-Rossen-2007-EmRev}
\bibinfo{author}{Ghysels, E.}, \bibinfo{author}{Arthur, S.},
  \bibinfo{author}{Rossen, V.}, \bibinfo{year}{2007}.
\newblock \bibinfo{title}{{MIDAS regressions: Further results and new
  directions}}.
\newblock \bibinfo{journal}{Econometr. Rev.} \bibinfo{volume}{26},
  \bibinfo{pages}{53--90}.
\newblock \DOIprefix\doi{10.1080/07474930600972467}.
\bibitem[{Ghysels et~al.(2004)Ghysels, Santa-Clara and
  Valkanov}]{Ghysels-Santa-Clara-Valkanov-2004}
\bibinfo{author}{Ghysels, E.}, \bibinfo{author}{Santa-Clara, P.},
  \bibinfo{author}{Valkanov, R.}, \bibinfo{year}{2004}.
\newblock \bibinfo{title}{{The MIDAS touch: Mixed data sampling regression
  models}}.
\newblock \bibinfo{type}{CIRANO Working Papers} \bibinfo{number}{2004s-20}.
  CIRANO.
\bibitem[{Hammoudeh et~al.(2014)Hammoudeh, Nguyen, Reboredo and
  Wen}]{Hammoudeh-Nguyen-Reboredo-Wen-2014-EMR}
\bibinfo{author}{Hammoudeh, S.}, \bibinfo{author}{Nguyen, D.K.},
  \bibinfo{author}{Reboredo, J.C.}, \bibinfo{author}{Wen, X.},
  \bibinfo{year}{2014}.
\newblock \bibinfo{title}{{Dependence of stock and commodity futures markets in
  China: Implications for portfolio investment}}.
\newblock \bibinfo{journal}{Emerg. Markets Rev.} \bibinfo{volume}{21},
  \bibinfo{pages}{183--200}.
\newblock \DOIprefix\doi{10.1016/j.ememar.2014.09.002}.
\bibitem[{Hasanov et~al.(2020)Hasanov, Shaiban and
  Freedi}]{Hasanov-Shaiban-Freedi-2020-EE}
\bibinfo{author}{Hasanov, A.S.}, \bibinfo{author}{Shaiban, M.S.},
  \bibinfo{author}{Freedi, A.}, \bibinfo{year}{2020}.
\newblock \bibinfo{title}{{Forecasting volatility in the petroleum futures
  markets: A re-examination and extension}}.
\newblock \bibinfo{journal}{Energy Econ.} \bibinfo{volume}{86},
  \bibinfo{pages}{104626}.
\newblock \DOIprefix\doi{10.1016/j.eneco.2019.104626}.
\bibitem[{Holmes and Otero(2019)}]{Holmes-Otero-2019-EE}
\bibinfo{author}{Holmes, M.J.}, \bibinfo{author}{Otero, J.},
  \bibinfo{year}{2019}.
\newblock \bibinfo{title}{{Re-examining the movements of crude oil spot and
  futures prices over time}}.
\newblock \bibinfo{journal}{Energy Econ.} \bibinfo{volume}{82},
  \bibinfo{pages}{224--236}.
\newblock \DOIprefix\doi{10.1016/j.eneco.2017.08.034}.
\bibitem[{Ji et~al.(2020)Ji, Zhang and Zhao}]{Ji-Zhang-Zhao-2020-IRFA}
\bibinfo{author}{Ji, Q.}, \bibinfo{author}{Zhang, D.}, \bibinfo{author}{Zhao,
  Y.}, \bibinfo{year}{2020}.
\newblock \bibinfo{title}{{Searching for safe-haven assets during the COVID-19
  pandemic}}.
\newblock \bibinfo{journal}{Int. Rev. Financ. Anal.} \bibinfo{volume}{71},
  \bibinfo{pages}{101526}.
\newblock \DOIprefix\doi{10.1016/j.irfa.2020.101526}.
\bibitem[{Jo{\"e}t et~al.(2017)Jo{\"e}t, Val{\'e}rie and
  Tovonony}]{Joet-Valerie-2017-EE}
\bibinfo{author}{Jo{\"e}t, M.}, \bibinfo{author}{Val{\'e}rie, M.},
  \bibinfo{author}{Tovonony, R.}, \bibinfo{year}{2017}.
\newblock \bibinfo{title}{{Does the volatility of commodity prices reflect
  macroeconomic uncertainty?}}
\newblock \bibinfo{journal}{Energy Econ.} \bibinfo{volume}{68},
  \bibinfo{pages}{313--326}.
\newblock \DOIprefix\doi{10.1016/j.eneco.2017.09.017}.
\bibitem[{Jones and Kaul(1996)}]{Jones-Kaul-1996-JF}
\bibinfo{author}{Jones, C.M.}, \bibinfo{author}{Kaul, G.},
  \bibinfo{year}{1996}.
\newblock \bibinfo{title}{{Oil and the stock markets}}.
\newblock \bibinfo{journal}{J. Financ.} \bibinfo{volume}{51},
  \bibinfo{pages}{463--491}.
\newblock \DOIprefix\doi{10.2307/2329368}.
\bibitem[{Jurado et~al.(2015)Jurado, Ludvigson and
  Ng}]{Jurado-Ludvigson-Ng-2015-AER}
\bibinfo{author}{Jurado, K.}, \bibinfo{author}{Ludvigson, S.C.},
  \bibinfo{author}{Ng, S.}, \bibinfo{year}{2015}.
\newblock \bibinfo{title}{{Measuring uncertainty}}.
\newblock \bibinfo{journal}{Amer. Econ. Rev.} \bibinfo{volume}{105},
  \bibinfo{pages}{1177--1216}.
\newblock \DOIprefix\doi{10.1257/aer.20131193}.
\bibitem[{Kang and Yoon(2013)}]{Kang-Yoon-2013-EE}
\bibinfo{author}{Kang, S.H.}, \bibinfo{author}{Yoon, S.M.},
  \bibinfo{year}{2013}.
\newblock \bibinfo{title}{{Modeling and forecasting the volatility of petroleum
  futures prices}}.
\newblock \bibinfo{journal}{Energy Econ.} \bibinfo{volume}{36},
  \bibinfo{pages}{354--362}.
\newblock \DOIprefix\doi{10.1016/j.eneco.2012.09.010}.
\bibitem[{Klein(2017)}]{Klein-2017-FRL}
\bibinfo{author}{Klein, T.}, \bibinfo{year}{2017}.
\newblock \bibinfo{title}{{Dynamic correlation of precious metals and
  flight-to-quality in developed markets}}.
\newblock \bibinfo{journal}{Financ. Res. Lett.} \bibinfo{volume}{23},
  \bibinfo{pages}{283--290}.
\newblock \DOIprefix\doi{10.1016/j.frl.2017.05.002}.
\bibitem[{Liu et~al.(2019)Liu, Pan, Yuan and
  Chen}]{Liu-Pan-Yuan-Chen-2019-Energy}
\bibinfo{author}{Liu, X.}, \bibinfo{author}{Pan, F.}, \bibinfo{author}{Yuan,
  L.}, \bibinfo{author}{Chen, Y.}, \bibinfo{year}{2019}.
\newblock \bibinfo{title}{{The dependence structure between crude oil futures
  prices and Chinese agricultural commodity futures prices: Measurement based
  on Markov-switching GRG copula}}.
\newblock \bibinfo{journal}{Energy} \bibinfo{volume}{182},
  \bibinfo{pages}{999--1012}.
\newblock \DOIprefix\doi{10.1016/j.energy.2019.06.071}.
\bibitem[{Liu et~al.(2018)Liu, Han and Yin}]{Liu-Han-Yin-2018-JFutM}
\bibinfo{author}{Liu, Y.}, \bibinfo{author}{Han, L.}, \bibinfo{author}{Yin,
  L.}, \bibinfo{year}{2018}.
\newblock \bibinfo{title}{{Does news uncertainty matter for commodity futures
  markets? Heterogeneity in energy and non-energy sectors}}.
\newblock \bibinfo{journal}{J. Fut. Markets} \bibinfo{volume}{38},
  \bibinfo{pages}{1246--1261}.
\newblock \DOIprefix\doi{10.1002/fut.21916}.
\bibitem[{Lucey et~al.(2017)Lucey, Sharma and
  Vigne}]{Lucey-Sharma-Vigne-2017-EconM}
\bibinfo{author}{Lucey, B.M.}, \bibinfo{author}{Sharma, S.S.},
  \bibinfo{author}{Vigne, S.A.}, \bibinfo{year}{2017}.
\newblock \bibinfo{title}{{Gold and inflation(s)–A time-varying
  relationship}}.
\newblock \bibinfo{journal}{Econ. Model.} \bibinfo{volume}{67},
  \bibinfo{pages}{88--101}.
\newblock \DOIprefix\doi{10.1016/j.econmod.2016.10.008}.
\bibitem[{Lv and Shan(2013)}]{Lv-Shan-2013-PA}
\bibinfo{author}{Lv, X.D.}, \bibinfo{author}{Shan, X.}, \bibinfo{year}{2013}.
\newblock \bibinfo{title}{{Modeling natural gas market volatility using GARCH
  with different distributions}}.
\newblock \bibinfo{journal}{Physica A} \bibinfo{volume}{392},
  \bibinfo{pages}{5685--5699}.
\newblock \DOIprefix\doi{10.1016/j.physa.2013.07.038}.
\bibitem[{Manela and Moreira(2017)}]{Manela-Moreira-2017-JFE}
\bibinfo{author}{Manela, A.}, \bibinfo{author}{Moreira, A.},
  \bibinfo{year}{2017}.
\newblock \bibinfo{title}{{News implied volatility and disaster concerns}}.
\newblock \bibinfo{journal}{J. Financ. Econ.} \bibinfo{volume}{123},
  \bibinfo{pages}{137--162}.
\newblock \DOIprefix\doi{10.1016/j.jfineco.2016.01.032}.
\bibitem[{Moore(2017)}]{Moore-2017-ER}
\bibinfo{author}{Moore, A.}, \bibinfo{year}{2017}.
\newblock \bibinfo{title}{{Measuring economic uncertainty and its effects}}.
\newblock \bibinfo{journal}{Econ. Rec.} \bibinfo{volume}{93},
  \bibinfo{pages}{550--575}.
\newblock \DOIprefix\doi{10.1111/1475-4932.12356}.
\bibitem[{Narayan et~al.(2010)Narayan, Narayan and
  Zheng}]{Narayan-Narayan-Zheng-2010-AEn}
\bibinfo{author}{Narayan, P.K.}, \bibinfo{author}{Narayan, S.},
  \bibinfo{author}{Zheng, X.}, \bibinfo{year}{2010}.
\newblock \bibinfo{title}{{Gold and oil futures markets: Are markets
  efficient?}}
\newblock \bibinfo{journal}{Appl. Energy} \bibinfo{volume}{87},
  \bibinfo{pages}{3299--3303}.
\newblock \DOIprefix\doi{10.1016/j.apenergy.2010.03.020}.
\bibitem[{Nguyen et~al.(2020)Nguyen, Sensoy, Sousa and
  Uddin}]{Nguyen-Sensoy-Sousa-Uddin-2020-EE}
\bibinfo{author}{Nguyen, D.K.}, \bibinfo{author}{Sensoy, A.},
  \bibinfo{author}{Sousa, R.M.}, \bibinfo{author}{Uddin, G.S.},
  \bibinfo{year}{2020}.
\newblock \bibinfo{title}{{U.S. equity and commodity futures markets: Hedging
  or financialization?}}
\newblock \bibinfo{journal}{Energy Econ.} \bibinfo{volume}{86},
  \bibinfo{pages}{104660}.
\newblock \DOIprefix\doi{10.1016/j.eneco.2019.104660}.
\bibitem[{Nguyen and Walther(2020)}]{Nguyen-Walther-2020-JFc}
\bibinfo{author}{Nguyen, D.K.}, \bibinfo{author}{Walther, T.},
  \bibinfo{year}{2020}.
\newblock \bibinfo{title}{{Modeling and forecasting commodity market volatility
  with long-term economic and financial variables}}.
\newblock \bibinfo{journal}{J. Forecast.} \bibinfo{volume}{39},
  \bibinfo{pages}{126--142}.
\newblock \DOIprefix\doi{10.1002/for.2617}.
\bibitem[{P\'astor and Veronesi(2012)}]{Pastor-Veronesi-2012-JF}
\bibinfo{author}{P\'astor, L.}, \bibinfo{author}{Veronesi, P.},
  \bibinfo{year}{2012}.
\newblock \bibinfo{title}{{Uncertainty about government policy and stock
  prices}}.
\newblock \bibinfo{journal}{J. Financ.} \bibinfo{volume}{67},
  \bibinfo{pages}{1219--1264}.
\newblock \DOIprefix\doi{10.1111/j.1540-6261.2012.01746.x}.
\bibitem[{P\'astor and Veronesi(2013)}]{Pastor-Veronesi-2013-JFE}
\bibinfo{author}{P\'astor, L.}, \bibinfo{author}{Veronesi, P.},
  \bibinfo{year}{2013}.
\newblock \bibinfo{title}{{Political uncertainty and risk premia}}.
\newblock \bibinfo{journal}{J. Financ. Econ.} \bibinfo{volume}{110},
  \bibinfo{pages}{520--545}.
\newblock \DOIprefix\doi{10.1016/j.jfineco.2013.08.007}.
\bibitem[{Sadorsky(1999)}]{Sadorsky-1999-EE}
\bibinfo{author}{Sadorsky, P.}, \bibinfo{year}{1999}.
\newblock \bibinfo{title}{{Oil price shocks and stock market activity}}.
\newblock \bibinfo{journal}{Energy Econ.} \bibinfo{volume}{21},
  \bibinfo{pages}{449--469}.
\newblock \DOIprefix\doi{10.1016/S0140-9883(99)00020-1}.
\bibitem[{Wang et~al.(2020)Wang, Shao and Kim}]{Wang-Shao-Kim-2020-CSF}
\bibinfo{author}{Wang, J.}, \bibinfo{author}{Shao, W.}, \bibinfo{author}{Kim,
  J.}, \bibinfo{year}{2020}.
\newblock \bibinfo{title}{{Analysis of the impact of COVID-19 on the
  correlations between crude oil and agricultural futures}}.
\newblock \bibinfo{journal}{Chaos Solitons Fractals} \bibinfo{volume}{136},
  \bibinfo{pages}{109896}.
\newblock \DOIprefix\doi{10.1016/j.chaos.2020.109896}.
\bibitem[{Wei et~al.(2017)Wei, Liu, Lai and Y.}]{Wei-Liu-Lai-Hu-2017-EE}
\bibinfo{author}{Wei, Y.}, \bibinfo{author}{Liu, J.}, \bibinfo{author}{Lai,
  X.}, \bibinfo{author}{Y., H.}, \bibinfo{year}{2017}.
\newblock \bibinfo{title}{{Which determinant is the most informative in
  forecasting crude oil market volatility: Fundamental, speculation, or
  uncertainty?}}
\newblock \bibinfo{journal}{Energy Econ.} \bibinfo{volume}{68},
  \bibinfo{pages}{141--150}.
\newblock \DOIprefix\doi{10.1016/j.eneco.2017.09.016}.
\bibitem[{Yan et~al.(2018)Yan, Irwin and Sanders}]{Yan-Irwin-Sanders-2018-EE}
\bibinfo{author}{Yan, L.}, \bibinfo{author}{Irwin, S.H.},
  \bibinfo{author}{Sanders, D.R.}, \bibinfo{year}{2018}.
\newblock \bibinfo{title}{{Mapping algorithms, agricultural futures, and the
  relationship between commodity investment flows and crude oil futures
  prices}}.
\newblock \bibinfo{journal}{Energy Econ.} \bibinfo{volume}{{72}},
  \bibinfo{pages}{486--504}.
\newblock \DOIprefix\doi{10.1016/j.eneco.2018.04.005}.
\bibitem[{Zhang et~al.(2019)Zhang, Ma and Wei}]{Zhang-Ma-Wei-2019-EE}
\bibinfo{author}{Zhang, Y.}, \bibinfo{author}{Ma, F.}, \bibinfo{author}{Wei,
  Y.}, \bibinfo{year}{2019}.
\newblock \bibinfo{title}{{Out-of-sample prediction of the oil futures market
  volatility: A comparison of new and traditional combination approaches}}.
\newblock \bibinfo{journal}{Energy Econ.} \bibinfo{volume}{81},
  \bibinfo{pages}{1109--1120}.
\newblock \DOIprefix\doi{10.1016/j.eneco.2019.05.018}.

\end{thebibliography}

\end{document}